\documentclass[10pt,twocolumn]{article}
\usepackage[margin=0.75in]{geometry}
\usepackage[T1]{fontenc}
\usepackage{tgtermes}
\usepackage{amsmath,amssymb}
\usepackage{booktabs}
\usepackage{multirow}
\usepackage{graphicx}
\usepackage{float}
\usepackage{tikz}
\usetikzlibrary{backgrounds}
\usepackage{pgfplots}
\usepgfplotslibrary{groupplots}
\pgfplotsset{compat=1.18}
\usepackage[
    backend=biber,
    style=ieee,
    sorting=none
]{biblatex}
\DeclareFieldFormat{eprint:wayback}{%
  Wayback Machine\addcolon\space
  \ifhyperref
    {\href{https://web.archive.org/web/#1/\thefield{eprintclass}}{\nolinkurl{https://web.archive.org/web/#1/\thefield{eprintclass}}}}
    {\nolinkurl{https://web.archive.org/web/#1/\thefield{eprintclass}}}%
}
\usepackage{fontawesome5}
\usepackage{microtype}
\usepackage{hyperref}

\hypersetup{
	pdftitle={8 Bits to End the Loudness War: Lossless Floating-Point Compression for Audio Streams},
	pdfauthor={Ali Heydari},
	colorlinks=true,
	linkcolor=black,
	citecolor=black,
	urlcolor=black
}

\definecolor{svgsign}{HTML}{C5FCFF}     
\definecolor{svgexp}{HTML}{9FFFAD}      
\definecolor{svgfrac}{HTML}{FFAEAD}     
\definecolor{svgteal}{HTML}{008080}     
\definecolor{svggreen}{HTML}{008000}    
\definecolor{svgmaroon}{HTML}{800000}   

\colorlet{grayexp}{svgexp!35!gray!45}

\colorlet{grayfrac}{svgfrac!40!gray!35}

\definecolor{wavegreen}{RGB}{41,155,108}

\pgfplotsset{
  colormap={bipolar_loudness}{
    rgb255(0cm)=(180,0,0)        
    rgb255(1.2cm)=(225,40,40)    
    rgb255(2.5cm)=(240,120,20)   
    rgb255(3.75cm)=(220,195,30)  
    rgb255(4.4cm)=(60,175,70)    
    rgb255(5.0cm)=(41,155,108)   
    rgb255(5.6cm)=(60,175,70)    
    rgb255(6.25cm)=(220,195,30)  
    rgb255(7.5cm)=(240,120,20)   
    rgb255(8.8cm)=(225,40,40)    
    rgb255(10.0cm)=(180,0,0)     
  }
}

\newcommand{\ceilinglines}{%
  \draw[red!80!black, dashed, line width=0.55pt, opacity=0.85] (axis cs:0, 1.0) -- (axis cs:232.2, 1.0);%
  \draw[red!80!black, dashed, line width=0.55pt, opacity=0.85] (axis cs:0, -1.0) -- (axis cs:232.2, -1.0);%
}

\title{\textbf{8 Bits to End the Loudness War: Lossless Floating-Point Compression for Audio Streams}}
\author{%
	\textbf{Ali Heydari}\\
	\small Department of Computer Engineering, Sharif University of Technology\\
	\small \href{mailto:ali.heydari.1381@sharif.edu}{\texttt{ali.heydari.1381@sharif.edu}}%
}
\date{%
	\vspace{0.5ex}%
	\normalsize\href{https://github.com/aliheidary1381/flac-with-float-coding}{\raisebox{-0.05ex}{\faGithub}\ \texttt{aliheidary1381/flac-with-float-coding}}\\[-1.5ex]%
}

\begin{document}

\maketitle

\begin{abstract}
	We present a native extension to the Free Lossless Audio Codec (FLAC) providing lossless compression for IEEE 754 single-precision (binary32) floating-point audio, alongside a survey and empirical benchmark of existing floating-point compression methods. While professional recording hardware is increasingly adopting 32-bit float converters and digital audio workstations process sound exclusively in floating-point math, consumer distribution remains constrained to fixed-point integer PCM. Through Dynamic Positional Alignment (DPA), our implementation achieves $\approx 44.1\%$ compression ratios on natural multi-stage ADC recordings---matching the storage footprint of standard 24-bit integer FLAC---and $\approx 60.2\%$ on complex digital workstation masters. By preserving dynamic peaks beyond 0\,dBFS, the format also allows listeners and audiophiles to undo clipping, providing a practical technical path away from the loudness war with virtually no storage overhead. We do not claim to introduce a fundamentally new compression paradigm; our approach is conceptually similar to the floating-point mechanism in WavPack (which remains undocumented without published literature). Instead, this paper compares existing methods, provides theoretical grounding for why they succeed or fail, benchmarks current codec implementations, and contributes a reference implementation of a proven method within the FLAC ecosystem. The source implementation code, the Internet-Draft specifying the changes to the standard, and the benchmarks are all available at the GitHub link.
\end{abstract}

\section{Introduction}

\subsection{The Floating-Point Pipeline and the Distribution Bottleneck}
Digital audio workflows are steadily transitioning toward end-to-end floating-point representation. Modern field recorders and audio interfaces have begun integrating 32-bit floating-point analog-to-digital converters (ADCs) \cite{FloatRecording}. By combining dual-stage converter topologies with staggered analog gains, these devices capture extended dynamic range without input-stage clipping. In parallel, Digital Audio Workstations (DAWs) have long performed mixing, processing, and summing in floating-point arithmetic. Within these production environments, signals that momentarily exceed 0\,dBFS do not lose information; intermediate peaks remain preserved across the IEEE 754 exponent field.

Despite this unified floating-point chain across capture and production, consumer distribution remains the missing link in the digital pipeline. Storage and release formats remain strictly constrained to fixed-point integer PCM (16-bit or 24-bit). In integer PCM, 0\,dBFS acts as an absolute digital boundary: sample values exceeding $\pm 1.0$ cannot be represented and are subject to hard digital clipping unless restrained by peak limiting. This forces engineers to compress or truncate dynamic peaks at the final export stage, breaking an otherwise continuous floating-point signal path.

\subsection[The Loudness War and Permanent Dynamic Loss]{The Loudness War and\\Permanent Dynamic Loss}
This rigid 0\,dBFS ceiling historically fueled the \textit{Loudness War} \cite{LoudnessWar}. In broadcast, streaming, and mobile listening environments, commercial incentives favored higher perceived acoustic loudness. Because fixed-point masters cannot exceed 0\,dBFS, mastering workflows increasingly turned to brickwall limiting and aggressive dynamic range compression to push average signal levels (RMS) closer to full scale.

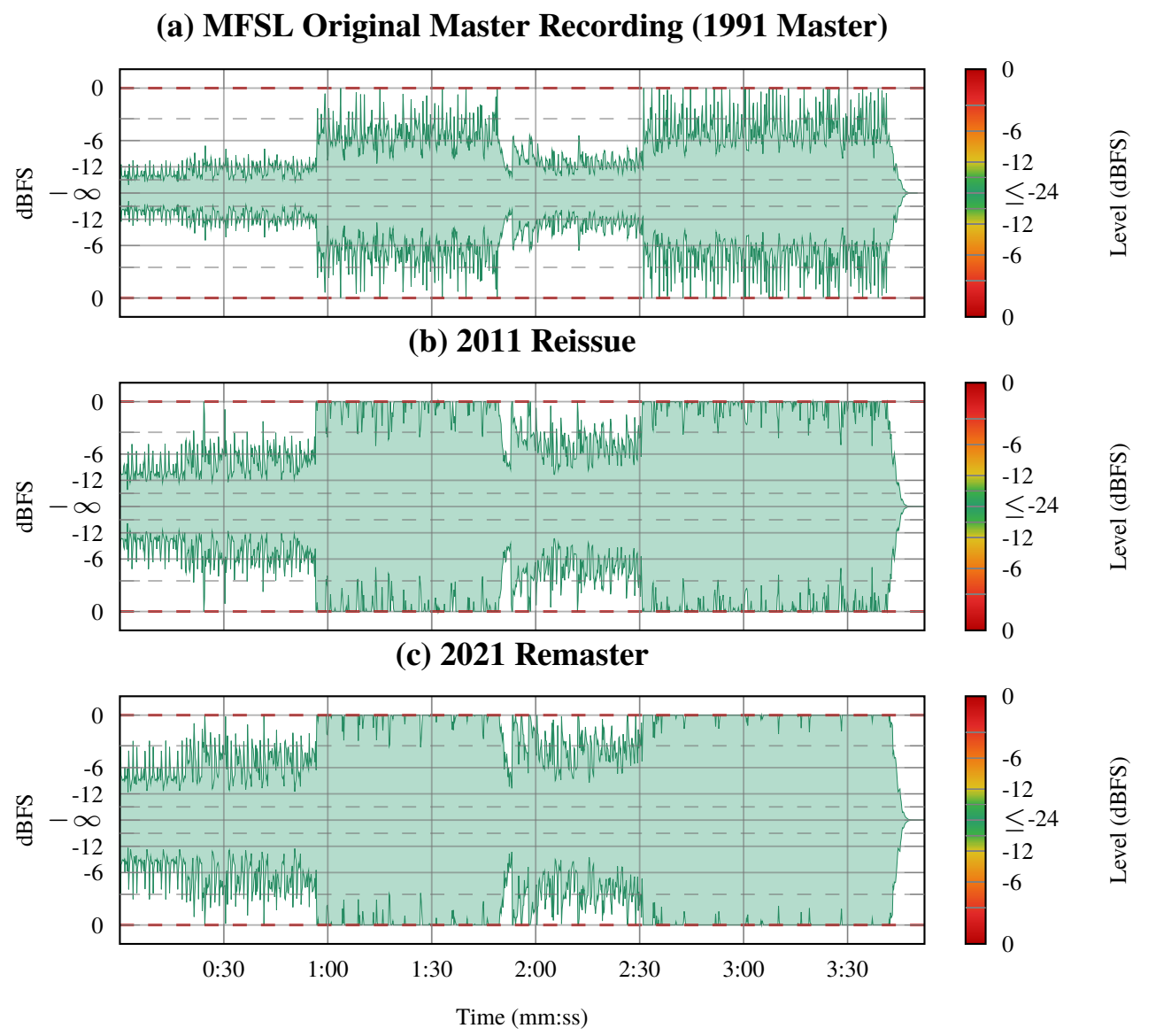
\begin{figure}[!ht]
	\centering
	\begin{tikzpicture}
		\begin{groupplot}[
				group style={
						group size=1 by 3,
						vertical sep=14pt,
						xlabels at=edge bottom,
						xticklabels at=edge bottom
					},
				scale only axis,
				axis on top,
				width=0.69\columnwidth,
				height=1.85cm,
				xmin=0, xmax=232.2,
				ymin=-1.18, ymax=1.18,
				grid=both,
				grid style={line width=0.2pt, draw=black!40, opacity=0.45},
				major grid style={line width=0.35pt, draw=black!55, opacity=0.55},
				ytick={-1.0, -0.5, -0.25, 0, 0.25, 0.5, 1.0},
				yticklabels={0, -6, -12, $-\infty$, -12, -6, 0},
				extra y ticks={-0.708, -0.126, 0.126, 0.708},
				extra y tick labels={},
				extra y tick style={
						grid=major,
						grid style={line width=0.2pt, draw=black!35, opacity=0.35, dashed}
					},
				xtick={30, 60, 90, 120, 150, 180, 210},
				xticklabels={0:30, 1:00, 1:30, 2:00, 2:30, 3:00, 3:30},
				yticklabel style={font=\tiny},
				xticklabel style={font=\tiny},
				ylabel={dBFS},
				ylabel style={font=\tiny, yshift=-4pt},
				colormap name=bipolar_loudness,
				point meta min=-1.0,
				point meta max=1.0,
				colorbar,
				colorbar style={
						ylabel={\tiny Level (dBFS)},
						ylabel style={font=\tiny},
						ytick={-1.0, -0.5, -0.25, 0, 0.25, 0.5, 1.0},
						yticklabels={0, -6, -12, $\le$-24, -12, -6, 0},
						yticklabel style={font=\tiny},
						width=0.15cm
					}
			]

			\nextgroupplot[
				title={\scriptsize (a) MFSL Original Master Recording (1991 Master)},
				title style={yshift=-4pt, font=\bfseries}
			]
			\addplot[fill=wavegreen!35, draw=none] table[x=t, y=y] {track1_poly.dat} -- cycle;
			\addplot[draw=wavegreen!90!black, line width=0.07pt] table[x=t, y=top] {track1.dat};
			\addplot[draw=wavegreen!90!black, line width=0.07pt] table[x=t, y=bottom] {track1.dat};
			\ceilinglines

			\nextgroupplot[
				title={\scriptsize (b) 2011 Reissue},
				title style={yshift=-4pt, font=\bfseries}
			]
			\addplot[fill=wavegreen!35, draw=none] table[x=t, y=y] {track2_poly.dat} -- cycle;
			\addplot[draw=wavegreen!90!black, line width=0.07pt] table[x=t, y=top] {track2.dat};
			\addplot[draw=wavegreen!90!black, line width=0.07pt] table[x=t, y=bottom] {track2.dat};
			\ceilinglines

			\nextgroupplot[
				title={\scriptsize (c) 2021 Remaster},
				title style={yshift=-4pt, font=\bfseries},
				xlabel={\tiny Time (mm:ss)}
			]
			\addplot[fill=wavegreen!35, draw=none] table[x=t, y=y] {track3_poly.dat} -- cycle;
			\addplot[draw=wavegreen!90!black, line width=0.07pt] table[x=t, y=top] {track3.dat};
			\addplot[draw=wavegreen!90!black, line width=0.07pt] table[x=t, y=bottom] {track3.dat};
			\ceilinglines

		\end{groupplot}
	\end{tikzpicture}
	\caption{Waveform and dynamic range comparison across three releases of Nirvana's \textit{Something In The Way} (1,280 sampled peaks per channel). Dashed red lines indicate the 0\,dBFS ($\pm 1.0$) digital ceiling. Notice the emergence of brickwall limiting along the $-0.3$\,dBFS ceiling during the choruses of the 2021 remaster compared to the dynamic headroom preserved in the MFSL release.}
	\label{fig:nirvana_clipping}
\end{figure}

\begin{table}[!ht]
	\centering
	\caption{Dynamic Range and Loudness Metrics Across Releases of Nirvana's \textit{Something In The Way}}
	\label{tab:loudness_comparison}
	\noindent\resizebox{\columnwidth}{!}{%
		\begin{tabular}{@{}lcccc@{}}
			\toprule
			\textbf{Release} & \textbf{Peak Level} & \textbf{RMS Level} & \textbf{Crest Factor} & $\mathbf{\Delta}$\textbf{RMS} \\
			                 & (dBFS)              & (dBFS)             & (dB)                  & vs. 1991                      \\ \midrule
			1991 MFSL Master & $-0.98$             & $-18.80$           & $17.82$               & \phantom{+}0.00\,dB           \\
			2011 Reissue     & $-0.30$             & $-13.15$           & $12.85$               & $+5.65$\,dB                   \\
			2021 Remaster    & $-0.33$             & $-11.64$           & $11.31$               & $+7.16$\,dB                   \\ \bottomrule
		\end{tabular}%
	}
\end{table}

While metadata-based volume normalization (e.g. ReplayGain \cite{ReplayGain}) was intended to mitigate perceptual loudness disparities, it is frequently bypassed, disabled, or inconsistently handled across playback hardware and consumer applications. This lack of reliability actively incentivizes mastering engineers to bake gain directly into the audio file itself---an inherently lossy practice that permanently destroys peak crest factor and waveform transients.

This dynamic trade-off is shown in Fig.~\ref{fig:nirvana_clipping} and Table~\ref{tab:loudness_comparison}, analyzing three releases of Nirvana's acoustic track \textit{Something In The Way} via FFmpeg's audio statistical filters \cite{ffmpeg}. In the 1991 Mobile Fidelity Sound Lab (MFSL) audiophile release, verses average around $-12$\,dBFS to $-18$\,dBFS, rising to peaks near $-1.0$\,dBFS during the chorus and maintaining a broad crest factor of $17.82$\,dB. In the 2021 remaster, the track's average RMS level is raised by $+7.16$\,dB, with peaks confined to a $-0.33$\,dBFS ceiling. Consequently, the crest factor drops by $6.51$\,dB, resulting in flat-topped peak envelopes during the loud passages.

Distributing audio in native IEEE 754 floating-point format provides a structural alternative. Peak transients that exceed 0\,dBFS during production remain mathematically preserved rather than clipped. Audiophiles and listeners who wish to experience the uncompressed crest factor can manually attenuate digital playback levels prior to digital-to-analog conversion, bringing the entire signal below 0\,dBFS and restoring the intended dynamic peaks without irreversible peak truncation.

The primary obstacle to this workflow has been the absence of native floating-point support in open lossless distribution formats. The Free Lossless Audio Codec (FLAC) \cite{flac} remains standard across high-resolution digital audio distribution, but has historically been restricted to integer PCM up to 32 bits. We implement native IEEE 754 single-precision compression within libFLAC, enabling float distribution without storage overhead.

\section{Methodology}
The encoding architecture presented in this section overcomes these distribution and dynamic compression challenges with zero storage overhead. Even where individual clipping artifacts or quantization losses might appear small in isolation, resolving them introduces no penalty in file size or transmission bandwidth. Our method is similar to WavPack \cite{WavPack}.

\subsection{IEEE 754 binary32 Structure}

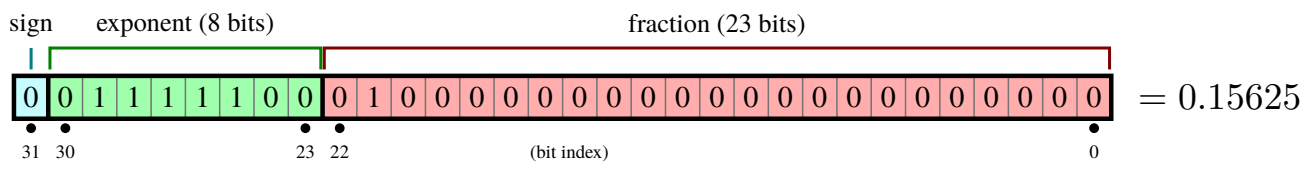
\begin{figure}[H]
	\centering
	\resizebox{\columnwidth}{!}{%
		\begin{tikzpicture}[x=0.32cm, y=0.40cm]
			\fill[svgsign] (0,0) rectangle (1,1);
			\fill[svgexp]  (1,0) rectangle (9,1);
			\fill[svgfrac] (9,0) rectangle (32,1);

			\node at (0.5, 0.5) {\footnotesize 0};

			\foreach \val [count=\i from 0] in {0,1,1,1,1,1,0,0} {
					\node at (1.5+\i, 0.5) {\footnotesize \val};
				}

			\foreach \val [count=\i from 0] in {0,1,0,0,0,0,0,0,0,0,0,0,0,0,0,0,0,0,0,0,0,0,0} {
					\node at (9.5+\i, 0.5) {\footnotesize \val};
				}

			\foreach \x in {1,...,31} {
					\draw[black!55, line width=0.4pt] (\x,0) -- (\x,1);
				}

			\draw[black, line width=1.2pt] (0,0) rectangle (32,1);
			\draw[black, line width=1.2pt] (1,0) -- (1,1);
			\draw[black, line width=1.2pt] (9,0) -- (9,1);

			\draw[svgteal, line width=0.8pt] (0.5, 1.15) -- (0.5, 1.65);
			\node[anchor=south] at (0.5, 1.6) {\scriptsize sign};

			\draw[svggreen, line width=0.8pt] (1.05, 1.15) -- (1.05, 1.65) -- (8.95, 1.65) -- (8.95, 1.15);
			\node[anchor=south] at (5, 1.6) {\scriptsize exponent (8 bits)};

			\draw[svgmaroon, line width=0.8pt] (9.05, 1.15) -- (9.05, 1.65) -- (31.95, 1.65) -- (31.95, 1.15);
			\node[anchor=south] at (20.5, 1.6) {\scriptsize fraction (23 bits)};

			\foreach \x/\label in {0.5/31, 1.5/30, 8.5/23, 9.5/22, 31.5/0} {
					\fill[black] (\x, -0.22) circle (1.3pt);
					\node[anchor=north] at (\x, -0.32) {\tiny \label};
				}
			\node[anchor=north] at (16.2, -0.32) {\tiny (bit index)};

			\node[anchor=west] at (32.4, 0.5) {\normalsize $= 0.15625$};
		\end{tikzpicture}%
	}
	\caption{Bit allocation of an IEEE 754 binary32 floating-point word (1-bit sign, 8-bit exponent, 23-bit significand fraction).}
	\label{fig:ieee_float}
\end{figure}

An IEEE 754 single-precision floating-point word comprises a 1-bit sign ($s$), an 8-bit biased exponent ($e$), and a 23-bit fraction ($m$), shown in Fig.~\ref{fig:ieee_float}. Defining the unbiased exponent as $e^* = e - 127$, the value evaluates to:
\begin{equation}
	v = (-1)^s \times 2^{e^*} \times \left(1 + \sum_{i=1}^{23} b_{23-i} 2^{-i}\right)
\end{equation}

Applying Linear Predictive Coding (LPC) directly to raw floating-point bit streams is ineffective. Sign-magnitude encoding and frequent exponent changes degrade sample-to-sample autocorrelation. In particular, transitions across power-of-two amplitude boundaries alter mantissa bits in step changes that integer predictors treat as high-frequency noise. However, natural audio signals generally vary smoothly, keeping the active bit span within a standard FLAC frame window ($N = 4096$ samples, $\approx 92.8$\,ms at 44.1\,kHz) well within manageable limits.

\subsection{Dynamic Positional Alignment (DPA) Mode}
In lossless audio compression, predictive coding cannot be formulated directly over floating-point representations due to a fundamental entropy explosion caused by precision dilation. Audio samples often possess sparse or compact mantissas (for example, an exact value of $1.0$ contains 23 trailing zero fraction bits). However, evaluating an LPC filter $\hat{x}_i = \sum_{k=1}^p a_k x_{i-k}$ with real-valued coefficients (such as $a_k \approx 1/3$) instantly populates all 23 fraction bits with non-zero arithmetic digits.

Even when a floating-point prediction is exceptionally accurate, differing from the ground truth by only a minimal epsilon near machine precision ($\approx 2^{-23}$), that fractional discrepancy expands the sample's active bit span by up to 23 bits. In an aligned integer register, this single bit of fractional error scatters non-zero bits across the entire lower register. Instead of yielding compact, near-zero discrete integer residuals ($0, \pm 1, \pm 2$) that Golomb-Rice coding compresses to 1--2 bits, the prediction error inflates into a massive 23-bit residual, destroying the Laplacian distribution required for entropy coding.

To prevent this precision dilation, predictive coding must operate over a discrete integer ring where predictor states are systematically discretized ($\hat{x}_i \in \mathbb{Z}$), requiring floating-point samples to first be aligned and mapped into discrete integer registers.

In binary positional notation, every normalized sample, $\pm 1.m \times 2^{e_i^*}$, has an explicit 24-bit mantissa~$M$. Its active bit boundary relative to $2^0$ is bounded by:
\begin{align}
	\text{pos}_{\max}(i) & = e_i^*                        \\
	\text{pos}_{\min}(i) & = e_i^* - 23 + \text{ctz}(M_i)
\end{align}
where $\text{ctz}(\cdot)$ denotes the count of trailing zeros. Across an audio frame of $N$ samples, the global active positions are:
\begin{equation}
	P_{\text{high}} = \max_i e_i^*, \quad P_{\text{low}} = \min_i(e_i^* - 23 + \text{ctz}(M_i))
\end{equation}

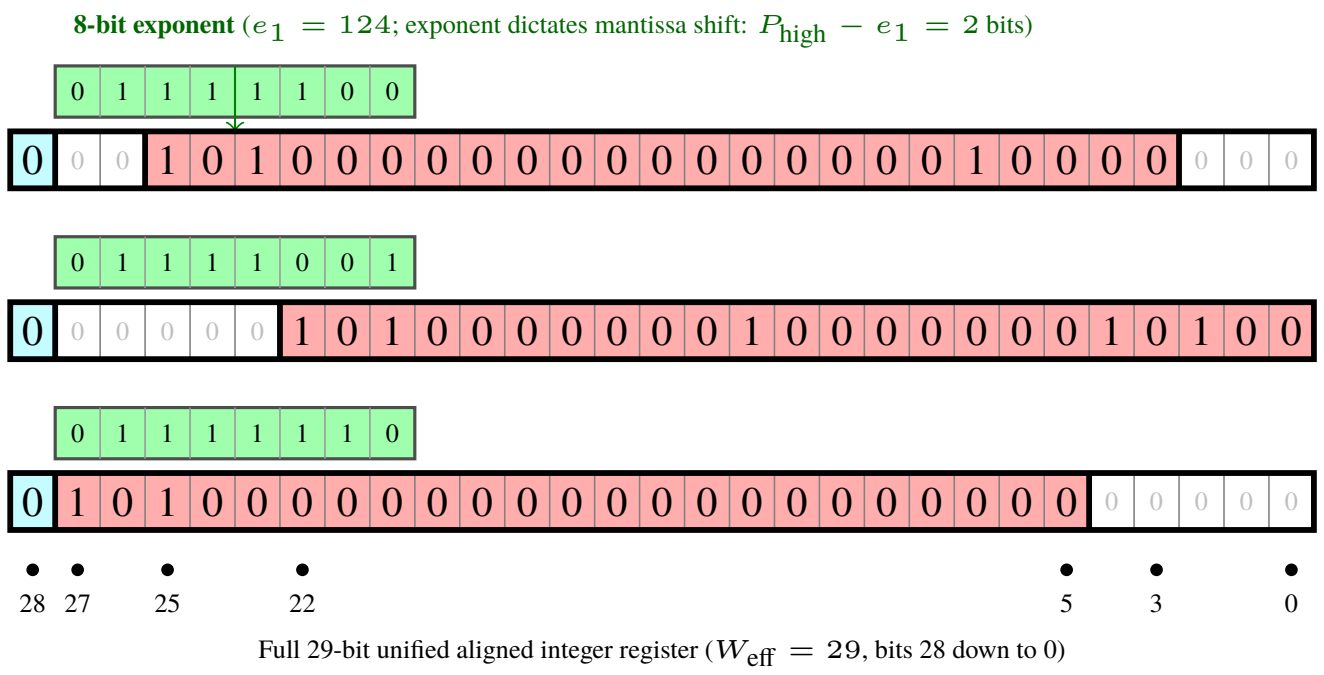
\begin{figure}[!ht]
	\centering
	\resizebox{\columnwidth}{!}{%
		\begin{tikzpicture}[x=0.30cm, y=0.38cm]


			\begin{scope}[on background layer]
				\fill[svgexp] (1, 7.35) rectangle (9, 8.25);
				\draw[black!70, line width=0.6pt] (1, 7.35) rectangle (9, 8.25);
				\foreach \x in {2,...,8} { \draw[black!40, line width=0.3pt] (\x, 7.35) -- (\x, 8.25); }
				\foreach \val [count=\i from 0] in {0,1,1,1,1,1,0,0} {
						\node[text=black] at (1.5+\i, 7.8) {\tiny \val};
					}
				\node[anchor=south west, text=green!40!black] at (1, 8.25) {\tiny \textbf{8-bit exponent} ($e_1 = 124$; exponent dictates mantissa shift: $P_{\text{high}} - e_1 = 2$ bits)};
				\draw[green!50!black, line width=0.4pt, ->] (5, 8.25) -- (5, 7.1);
			\end{scope}

			\fill[svgsign] (0, 6.1) rectangle (1, 7.1);
			\node at (0.5, 6.6) {\footnotesize 0};

			\fill[white] (1, 6.1) rectangle (3, 7.1);
			\node[text=black!25] at (1.5, 6.6) {\tiny 0};
			\node[text=black!25] at (2.5, 6.6) {\tiny 0};

			\fill[svgfrac] (3, 6.1) rectangle (26, 7.1);
			\node[text=black] at (3.5, 6.6)  {\footnotesize 1};
			\node[text=black] at (4.5, 6.6)  {\footnotesize 0};
			\node[text=black] at (5.5, 6.6)  {\footnotesize 1};
			\foreach \x in {6.5, 7.5, ..., 20.5} { \node[text=black] at (\x, 6.6) {\footnotesize 0}; }
			\node[text=black] at (21.5, 6.6) {\footnotesize 1};
			\node[text=black] at (22.5, 6.6) {\footnotesize 0};
			\node[text=black] at (23.5, 6.6) {\footnotesize 0};
			\node[text=black] at (24.5, 6.6) {\footnotesize 0};
			\node[text=black] at (25.5, 6.6) {\footnotesize 0};

			\fill[white] (26, 6.1) rectangle (29, 7.1);
			\node[text=black!25] at (26.5, 6.6) {\tiny 0};
			\node[text=black!25] at (27.5, 6.6) {\tiny 0};
			\node[text=black!25] at (28.5, 6.6) {\tiny 0};

			\foreach \x in {1,...,28} { \draw[black!50, line width=0.3pt] (\x, 6.1) -- (\x, 7.1); }
			\draw[black, line width=1.1pt] (0, 6.1) rectangle (29, 7.1);
			\draw[black, line width=1.1pt] (1, 6.1) -- (1, 7.1);
			\draw[black, line width=1.1pt] (3, 6.1) -- (3, 7.1);
			\draw[black, line width=1.1pt] (26, 6.1) -- (26, 7.1);


			\begin{scope}[on background layer]
				\fill[svgexp] (1, 4.35) rectangle (9, 5.25);
				\draw[black!70, line width=0.6pt] (1, 4.35) rectangle (9, 5.25);
				\foreach \x in {2,...,8} { \draw[black!40, line width=0.3pt] (\x, 4.35) -- (\x, 5.25); }
				\foreach \val [count=\i from 0] in {0,1,1,1,1,0,0,1} {
						\node[text=black] at (1.5+\i, 4.8) {\tiny \val};
					}
			\end{scope}

			\fill[svgsign] (0, 3.1) rectangle (1, 4.1);
			\node at (0.5, 3.6) {\footnotesize 0};

			\fill[white] (1, 3.1) rectangle (6, 4.1);
			\foreach \x in {1.5, 2.5, ..., 5.5} { \node[text=black!25] at (\x, 3.6) {\tiny 0}; }

			\fill[svgfrac] (6, 3.1) rectangle (29, 4.1);
			\node[text=black] at (6.5, 3.6) {\footnotesize 1};
			\node[text=black] at (7.5, 3.6) {\footnotesize 0};
			\node[text=black] at (8.5, 3.6) {\footnotesize 1};
			\foreach \x in {9.5, 10.5, ..., 15.5} { \node[text=black] at (\x, 3.6) {\footnotesize 0}; }
			\node[text=black] at (16.5, 3.6) {\footnotesize 1};
			\node[text=black] at (17.5, 3.6) {\footnotesize 0};
			\foreach \x in {18.5, 19.5, ..., 23.5} { \node[text=black] at (\x, 3.6) {\footnotesize 0}; }
			\node[text=black] at (24.5, 3.6) {\footnotesize 1};
			\node[text=black] at (25.5, 3.6) {\footnotesize 0};
			\node[text=black] at (26.5, 3.6) {\footnotesize 1};
			\node[text=black] at (27.5, 3.6) {\footnotesize 0};
			\node[text=black] at (28.5, 3.6) {\footnotesize 0};

			\foreach \x in {1,...,28} { \draw[black!50, line width=0.3pt] (\x, 3.1) -- (\x, 4.1); }
			\draw[black, line width=1.1pt] (0, 3.1) rectangle (29, 4.1);
			\draw[black, line width=1.1pt] (1, 3.1) -- (1, 4.1);
			\draw[black, line width=1.1pt] (6, 3.1) -- (6, 4.1);


			\begin{scope}[on background layer]
				\fill[svgexp] (1, 1.35) rectangle (9, 2.25);
				\draw[black!70, line width=0.6pt] (1, 1.35) rectangle (9, 2.25);
				\foreach \x in {2,...,8} { \draw[black!40, line width=0.3pt] (\x, 1.35) -- (\x, 2.25); }
				\foreach \val [count=\i from 0] in {0,1,1,1,1,1,1,0} {
						\node[text=black] at (1.5+\i, 1.8) {\tiny \val};
					}
			\end{scope}

			\fill[svgsign] (0, 0.1) rectangle (1, 1.1);
			\node at (0.5, 0.6) {\footnotesize 0};

			\fill[svgfrac] (1, 0.1) rectangle (24, 1.1);
			\node[text=black] at (1.5, 0.6) {\footnotesize 1};
			\node[text=black] at (2.5, 0.6) {\footnotesize 0};
			\node[text=black] at (3.5, 0.6) {\footnotesize 1};
			\foreach \x in {4.5, 5.5, ..., 23.5} { \node[text=black] at (\x, 0.6) {\footnotesize 0}; }

			\fill[white] (24, 0.1) rectangle (29, 1.1);
			\foreach \x in {24.5, 25.5, ..., 28.5} { \node[text=black!25] at (\x, 0.6) {\tiny 0}; }

			\foreach \x in {1,...,28} { \draw[black!50, line width=0.3pt] (\x, 0.1) -- (\x, 1.1); }
			\draw[black, line width=1.1pt] (0, 0.1) rectangle (29, 1.1);
			\draw[black, line width=1.1pt] (1, 0.1) -- (1, 1.1);
			\draw[black, line width=1.1pt] (24, 0.1) -- (24, 1.1);

			\foreach \x/\label in {0.5/28, 1.5/27, 3.5/25, 6.5/22, 23.5/5, 25.5/3, 28.5/0} {
					\fill[black] (\x, -0.6) circle (1.2pt);
					\node[anchor=north] at (\x, -0.7) {\tiny \label};
				}
			\node[anchor=north] at (14.5, -1.5) {\tiny Full 29-bit unified aligned integer register ($W_{\text{eff}} = 29$, bits 28 down to 0)};

		\end{tikzpicture}%
	}
	\caption{Dynamic Positional Alignment (DPA) mapping single-precision IEEE 754 samples into a unified 29-bit integer register. The per-sample exponent dictates the alignment shift relative to the block maximum ($P_{\text{high}}$). Coral cells delineate the active 24-bit significand, while white cells represent inactive headroom and trailing zero padding.}
	\label{fig:unified_alignment}
\end{figure}

However, applying LPC directly to this unified aligned integer register introduces an information-theoretic conflict. While both regions of the word represent the amplitude-modulated signal, real-world acoustic recordings are strictly bounded by a physical Effective Number of Bits (ENOB) dictated by the ambient background noise of the recording environment. Even world-class studios certified to Noise Criterion NC-15 or NC-20 exhibit an ambient noise floor of $\approx 15$--$25$\,dBSPL. Against typical musical peaks ($\approx 115$--$125$\,dBSPL), the physical space alone caps the usable dynamic range to under $100$\,dB ($\approx 16.6$ bits). Below this acoustic threshold, recorded data consists entirely of uncorrelated environmental air motion and room noise, strictly limiting coherent, auto-correlated acoustic resolution to $\approx 16$ bits.

As a result, the aligned integer behaves as an alloy of two distinct physical regimes:
\begin{enumerate}
	\item \textbf{Upper Bits (Physical ENOB):} Corresponds to the coherent acoustic signal rising above the room's ambient noise floor. Because musical instruments and acoustic vibrations evolve smoothly over time, this region exhibits strong sample-to-sample autocorrelation. Linear prediction models this physical motion with high fidelity, yielding prediction residuals that strictly follow a zero-mean Laplacian (discrete two-sided geometric) distribution, ideally suited for Golomb-Rice entropy coding.
	\item \textbf{Lower Bits (Noise Floor and Rounding Arithmetic):} Corresponds to the sub-ENOB region---governed entirely by ambient room background noise in physical captures, or uniform floating-point rounding arithmetic in DAW production workflows. Although this noise floor is envelope-modulated across block exponent scales, its instantaneous bit values are uncorrelated and identically distributed ($p(b_j = 0) \approx p(b_j = 1) \approx 0.5$), behaving as pure white noise.
\end{enumerate}

When this incompressible uniform noise floor is passed into a partitioned Rice coder alongside the upper bits, the Rice parameter $k$ inflates dramatically across entire subframe partitions to prevent unary blowout, penalizing the coding efficiency of the correlated upper bits.

To resolve this, DPA employs a dynamic 2-way split that routes each distribution to its mathematically optimal entropy coder, parameterized by the upper subframe bit depth $W_{\text{upper}}$ (default 24 bits, stored in the frame header) and the lower subframe bit depth $W_{\text{lower}}$ ($0 \le W_{\text{lower}} \le 31$). The total effective alignment width and corresponding base exponent position are given by:
\begin{align}
	W_{\text{eff}}  & = W_{\text{upper}} + W_{\text{lower}}    \\
	P_{\text{base}} & = P_{\text{high}} - (W_{\text{eff}} - 2)
\end{align}

Each logical audio channel is segregated into two coupled subframes:
\begin{enumerate}
	\item \textbf{Upper Subframe ($W_{\text{upper}}$ bits):} Contains the correlated acoustic waveform:
	      \begin{equation}
		      x_{\text{upper}}[i] = x_i \gg W_{\text{lower}}
	      \end{equation}
	      This subframe is processed through the standard FLAC LPC and Rice entropy coding pipeline, where the Laplacian predictor residuals are compressed optimally.
	\item \textbf{Lower Subframe ($W_{\text{lower}}$ bits):} Contains the uncorrelated fractional noise:
	      \begin{equation}
		      x_{\text{lower}}[i] = x_i \ \& \ ((1 \ll W_{\text{lower}}) - 1)
	      \end{equation}
	      Because uniform noise cannot be compressed, this subframe bypasses Rice coding entirely and is serialized using a dedicated \texttt{PACKED} subframe type that writes raw, uncompressed variable-width bits. Capping the lower bit depth to at most 31 bits ($W_{\text{lower}} \le 31$) maintains a bounded lower subframe, as samples with dynamic spans exceeding this range can be represented in 32 bits as raw floats.
\end{enumerate}

During sample alignment, when facing exact zeros ($\pm 0.0\text{f}$), subnormal values, non-finite values ($\pm\infty$, NaNs), or values whose magnitude falls below the active upper window---the encoder sets $x_{\text{upper}}[i] = 0$ in the upper subframe and writes the entire 32-bit raw IEEE 754 floating-point pattern directly into the packed subframe stream as an uncompressed 32-bit literal.

At decode time, the decoder reconstructs the alignment parameters symmetrically:
\begin{align}
	W_{\text{eff}}  & = W_{\text{upper}} + W_{\text{lower}}    \\
	P_{\text{base}} & = P_{\text{high}} - (W_{\text{eff}} - 2)
\end{align}
For each sample:
\begin{itemize}
	\item If $x_{\text{upper}}[i] \ne 0$: The decoder reads $x_{\text{lower}}[i]$, reconstructs the aligned integer via bitwise reassembly:
	      \begin{equation}
		      x_i = (x_{\text{upper}}[i] \ll W_{\text{lower}}) \mid x_{\text{lower}}[i]
	      \end{equation}
	      and recovers the original floating-point exponent directly from $P_{\text{base}}$ and the leading bit position with zero overhead.
	\item If $x_{\text{upper}}[i] = 0$: The decoder reads 32 raw bits from the packed subframe bitstream and emits the float directly.
\end{itemize}

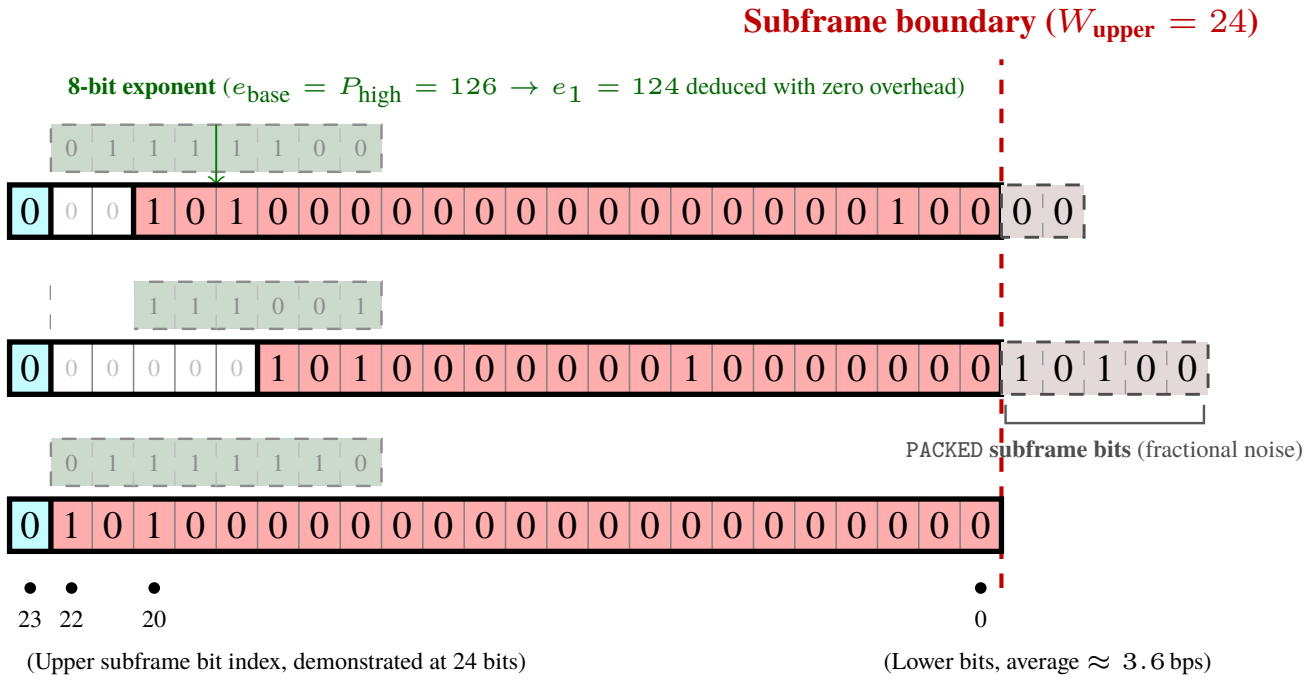
\begin{figure}[!ht]
	\centering
	\resizebox{\columnwidth}{!}{%
		\begin{tikzpicture}[x=0.30cm, y=0.38cm]

			\draw[red!75!black, dashed, line width=0.85pt] (24, -0.6) -- (24, 9.5);
			\node[anchor=south, text=red!75!black] at (24, 9.5) {\scriptsize \textbf{Subframe boundary ($W_{\text{upper}} = 24$)}};


			\begin{scope}[on background layer]
				\fill[grayexp] (1, 7.35) rectangle (9, 8.25);
				\draw[black!45, dashed, line width=0.5pt] (1, 7.35) rectangle (9, 8.25);
				\foreach \x in {2,...,8} { \draw[black!25, dashed, line width=0.3pt] (\x, 7.35) -- (\x, 8.25); }
				\foreach \val [count=\i from 0] in {0,1,1,1,1,1,0,0} {
						\node[text=black!45] at (1.5+\i, 7.8) {\tiny \val};
					}
				\node[anchor=south west, text=green!40!black] at (1, 8.25) {\tiny \textbf{8-bit exponent} ($e_{\text{base}} = P_{\text{high}} = 126 \to e_1 = 124$ deduced with zero overhead)};
				\draw[green!50!black, line width=0.4pt, ->] (5, 8.25) -- (5, 7.1);
			\end{scope}

			\fill[svgsign] (0, 6.1) rectangle (1, 7.1);
			\node at (0.5, 6.6) {\footnotesize 0};

			\fill[white] (1, 3) rectangle (3, 7.1);
			\node[text=black!25] at (1.5, 6.6) {\tiny 0};
			\node[text=black!25] at (2.5, 6.6) {\tiny 0};

			\fill[svgfrac] (3, 6.1) rectangle (24, 7.1);
			\node[text=black] at (3.5, 6.6)  {\footnotesize 1};
			\node[text=black] at (4.5, 6.6)  {\footnotesize 0};
			\node[text=black] at (5.5, 6.6)  {\footnotesize 1};
			\foreach \x in {6.5, 7.5, ..., 20.5} { \node[text=black] at (\x, 6.6) {\footnotesize 0}; }
			\node[text=black] at (21.5, 6.6) {\footnotesize 1};
			\node[text=black] at (22.5, 6.6) {\footnotesize 0};
			\node[text=black] at (23.5, 6.6) {\footnotesize 0};

			\foreach \x in {1,...,23} { \draw[black!50, line width=0.3pt] (\x, 6.1) -- (\x, 7.1); }
			\draw[black, line width=1.1pt] (0, 6.1) rectangle (24, 7.1);
			\draw[black, line width=1.1pt] (1, 6.1) -- (1, 7.1);
			\draw[black, line width=1.1pt] (3, 6.1) -- (3, 7.1);

			\fill[grayfrac] (24, 6.1) rectangle (26, 7.1);
			\node[text=black] at (24.5, 6.6) {\footnotesize 0};
			\node[text=black] at (25.5, 6.6) {\footnotesize 0};
			\draw[black!40, dashed, line width=0.3pt] (25, 6.1) -- (25, 7.1);
			\draw[black!70, dashed, line width=0.6pt] (24, 6.1) rectangle (26, 7.1);


			\begin{scope}[on background layer]
				\fill[grayexp] (1, 4.35) rectangle (9, 5.25);
				\draw[black!45, dashed, line width=0.5pt] (1, 4.35) rectangle (9, 5.25);
				\foreach \x in {2,...,8} { \draw[black!25, dashed, line width=0.3pt] (\x, 4.35) -- (\x, 5.25); }
				\foreach \val [count=\i from 0] in {0,1,1,1,1,0,0,1} {
						\node[text=black!45] at (1.5+\i, 4.8) {\tiny \val};
					}
			\end{scope}

			\fill[svgsign] (0, 3.1) rectangle (1, 4.1);
			\node at (0.5, 3.6) {\footnotesize 0};

			\fill[white] (1, 3.1) rectangle (6, 4.1);
			\foreach \x in {1.5, 2.5, ..., 5.5} { \node[text=black!25] at (\x, 3.6) {\tiny 0}; }

			\fill[svgfrac] (6, 3.1) rectangle (24, 4.1);
			\node[text=black] at (6.5, 3.6) {\footnotesize 1};
			\node[text=black] at (7.5, 3.6) {\footnotesize 0};
			\node[text=black] at (8.5, 3.6) {\footnotesize 1};
			\foreach \x in {9.5, 10.5, ..., 15.5} { \node[text=black] at (\x, 3.6) {\footnotesize 0}; }
			\node[text=black] at (16.5, 3.6) {\footnotesize 1};
			\node[text=black] at (17.5, 3.6) {\footnotesize 0};
			\foreach \x in {18.5, 19.5, ..., 23.5} { \node[text=black] at (\x, 3.6) {\footnotesize 0}; }

			\foreach \x in {1,...,23} { \draw[black!50, line width=0.3pt] (\x, 3.1) -- (\x, 4.1); }
			\draw[black, line width=1.1pt] (0, 3.1) rectangle (24, 4.1);
			\draw[black, line width=1.1pt] (1, 3.1) -- (1, 4.1);
			\draw[black, line width=1.1pt] (6, 3.1) -- (6, 4.1);

			\fill[grayfrac] (24, 3.1) rectangle (29, 4.1);
			\node[text=black] at (24.5, 3.6) {\footnotesize 1};
			\node[text=black] at (25.5, 3.6) {\footnotesize 0};
			\node[text=black] at (26.5, 3.6) {\footnotesize 1};
			\node[text=black] at (27.5, 3.6) {\footnotesize 0};
			\node[text=black] at (28.5, 3.6) {\footnotesize 0};

			\foreach \x in {25, 26, 27, 28} {
					\draw[black!40, dashed, line width=0.3pt] (\x, 3.1) -- (\x, 4.1);
				}
			\draw[black!70, dashed, line width=0.6pt] (24, 3.1) rectangle (29, 4.1);

			\draw[black!65, line width=0.5pt] (24.1, 2.9) -- (24.1, 2.55) -- (28.9, 2.55) -- (28.9, 2.9);
			\node[anchor=north, text=black!70] at (26.5, 2.55) {\tiny \textbf{\texttt{PACKED} subframe bits} (fractional noise)};


			\begin{scope}[on background layer]
				\fill[grayexp] (1, 1.35) rectangle (9, 2.25);
				\draw[black!45, dashed, line width=0.5pt] (1, 1.35) rectangle (9, 2.25);
				\foreach \x in {2,...,8} { \draw[black!25, dashed, line width=0.3pt] (\x, 1.35) -- (\x, 2.25); }
				\foreach \val [count=\i from 0] in {0,1,1,1,1,1,1,0} {
						\node[text=black!45] at (1.5+\i, 1.8) {\tiny \val};
					}
			\end{scope}

			\fill[svgsign] (0, 0.1) rectangle (1, 1.1);
			\node at (0.5, 0.6) {\footnotesize 0};

			\fill[svgfrac] (1, 0.1) rectangle (24, 1.1);
			\node[text=black] at (1.5, 0.6) {\footnotesize 1};
			\node[text=black] at (2.5, 0.6) {\footnotesize 0};
			\node[text=black] at (3.5, 0.6) {\footnotesize 1};
			\foreach \x in {4.5, 5.5, ..., 23.5} { \node[text=black] at (\x, 0.6) {\footnotesize 0}; }

			\foreach \x in {1,...,23} { \draw[black!50, line width=0.3pt] (\x, 0.1) -- (\x, 1.1); }
			\draw[black, line width=1.1pt] (0, 0.1) rectangle (24, 1.1);
			\draw[black, line width=1.1pt] (1, 0.1) -- (1, 1.1);

			\foreach \x/\label in {0.5/23, 1.5/22, 3.5/20, 23.5/0} {
					\fill[black] (\x, -0.6) circle (1.2pt);
					\node[anchor=north] at (\x, -0.7) {\tiny \label};
				}
			\node[anchor=north west] at (0.0, -1.5) {\tiny (Upper subframe bit index, demonstrated at 24 bits)};
			\node[anchor=north east] at (29.5, -1.5) {\tiny (Lower bits, average $\approx 3.6$\,bps)};

		\end{tikzpicture}%
	}
	\caption{Dynamic Positional Alignment (DPA) 2-way split between the upper LPC subframe and lower \texttt{PACKED} subframe (upper subframe width demonstrated at 24 bits).}
	\label{fig:subframe_alignment}
\end{figure}

As illustrated in Fig.~\ref{fig:subframe_alignment}, this dynamic positional alignment unifies the exponent trajectory and organizes samples relative to the subframe boundary.

\subsection{Frame Mode Selection and Adaptive Fallback}
Operating modes are selected on a per-frame basis. A frame defaults to DPA mode. Under pathological dynamic spans or extreme frame-level dynamic range, the auxiliary \texttt{PACKED} subframe may theoretically expand to store full 32-bit uncompressed floating-point literals with no compression gain. When an audio segment contains extreme dynamic divergence, numerical outliers, or high concentrations of values that drop below the active upper window, the overhead of transmitting uncompressed literal payloads alongside upper subframe metadata can cause negative compression (expanding the file beyond the raw stream). To prevent this, the codec incorporates an adaptive fallback mechanism reserved for:
\begin{itemize}
	\item \textbf{Efficiency Reversion:} If the DPA compressed frame size fails to achieve at least a $25\%$ savings (i.e., the compressed payload is $>75\%$ of the raw uncompressed bits), the encoder automatically discards the DPA pass and reverts to Fallback mode for that frame.
\end{itemize}

In fallback mode, each audio channel is split into two independent subframes (4 subframes for stereo):
\begin{enumerate}
	\item \textbf{Exponent Subframe (8 bits):} Centered using a $-128$ DC offset to remove step discontinuities, establishing a continuous exponent trajectory that facilitates inter-channel decorrelation and LPC filtering.
	\item \textbf{Significand Subframe (24 bits):} The sign and fraction fields are mapped from sign-magnitude to signed two's complement, mapping $-0.0\text{f}$ uniquely to $-8{,}388{,}608$ and providing a continuous signed manifold around zero-crossings for linear prediction.
\end{enumerate}

\section{Performance and Implications}

In the comparative figures throughout this section, benchmarks are presented exclusively for Tier~3, 24-bit, $>48$\,kHz real-world streams. This focus reflects our primary target scenario: consumer delivery of high-resolution (Hi-Res) music. Our core empirical finding is that such high-resolution music can be losslessly transitioned to 32-bit floating-point representations with zero additional storage footprint compared to legacy 24-bit integer PCM. Readers interested in comprehensive benchmarking across all tiers, bit depths, sampling strata, and synthetic stress signals are referred to Appendix~\ref{sec:benchmarks}.

Across both capture and master simulations, FLAC exhibits a 3\% to 4\% compression lag relative to the best-performing codecs (WavPack, MPEG-4 ALS, and OptimFROG). Crucially, this margin is entirely expected: standard FLAC displays the exact same 3\% to 4\% gap when compressing conventional integer PCM streams due to its lightweight linear prediction engine and simple Rice entropy coding compared to high-complexity asymmetric numeral systems or advanced context models. The underlying mechanisms for floating-point segregation perform identically across codecs (with the sole exception of Monkey's Audio, which fails to isolate mantissa noise). Most importantly, compared to conventional 24-bit integer FLAC files, FLAC floating-point files still maintain a minimal storage footprint.

\subsection{Performance on Raw ADC Captures}
On live recordings captured via multi-stage 32-bit float ADCs, DPA mode achieves an exceptional $\approx 44.1\%$ compression ratio, matching the storage footprint of standard 24-bit integer FLAC. This compression performance arises directly from the underlying physical capture topology: raw audio is recorded by physical ADCs that operate with an effective resolution of 24 bits. Although the signal is scaled into floating-point format via gain staging, the active bit span within each individual frame remains bounded by 24 bits. The only transition occurs when the recording hardware switches between converter stages, which happens quite infrequently across standard frame windows.

\begin{figure}[!ht]
	\centering
	\begin{tikzpicture}
		\begin{axis}[
				ybar,
				width=0.98\columnwidth,
				height=4.8cm,
				bar width=0.55cm,
				ymin=0, ymax=105,
				ytick={0, 20, 40, 60, 80, 100},
				yticklabels={0\%, 20\%, 40\%, 60\%, 80\%, 100\%},
				ylabel={Compression Ratio ($\downarrow$ lower is better)},
				ylabel style={font=\scriptsize},
				symbolic x coords={FLAC, WavPack, MPEG-4 ALS, OptimFROG, Monkey's Audio},
				xtick=data,
				nodes near coords={\pgfmathprintnumber\pgfplotspointmeta\%},
				nodes near coords style={font=\footnotesize, /pgf/number format/fixed, /pgf/number format/precision=1},
				grid=both,
				grid style={line width=0.2pt, draw=black!40, opacity=0.45},
				major grid style={line width=0.35pt, draw=black!55, opacity=0.55},
				enlarge x limits=0.18,
				x tick label style={font=\footnotesize, rotate=30, anchor=east}
			]
			\draw[->, >=stealth, line width=0.55pt, draw=black!65] (rel axis cs:0.04, 0.93) -- (rel axis cs:0.04, 0.77)
			node[midway, right=1.5pt, font=\tiny, text=black!75] {Lower is better};
			\addplot[fill=wavegreen!65, draw=wavegreen!90!black] coordinates {
					(FLAC, 44.11)
					(WavPack, 40.87)
					(MPEG-4 ALS, 40.36)
					(OptimFROG, 40.70)
					(Monkey's Audio, 89.81)
				};
		\end{axis}
	\end{tikzpicture}%
	\caption{Compression ratio across codecs under Dual-Stage ADC simulation on high-resolution musical material (Tier~3, 24-bit, $>48$\,kHz).}
	\label{fig:chart_adc}
\end{figure}
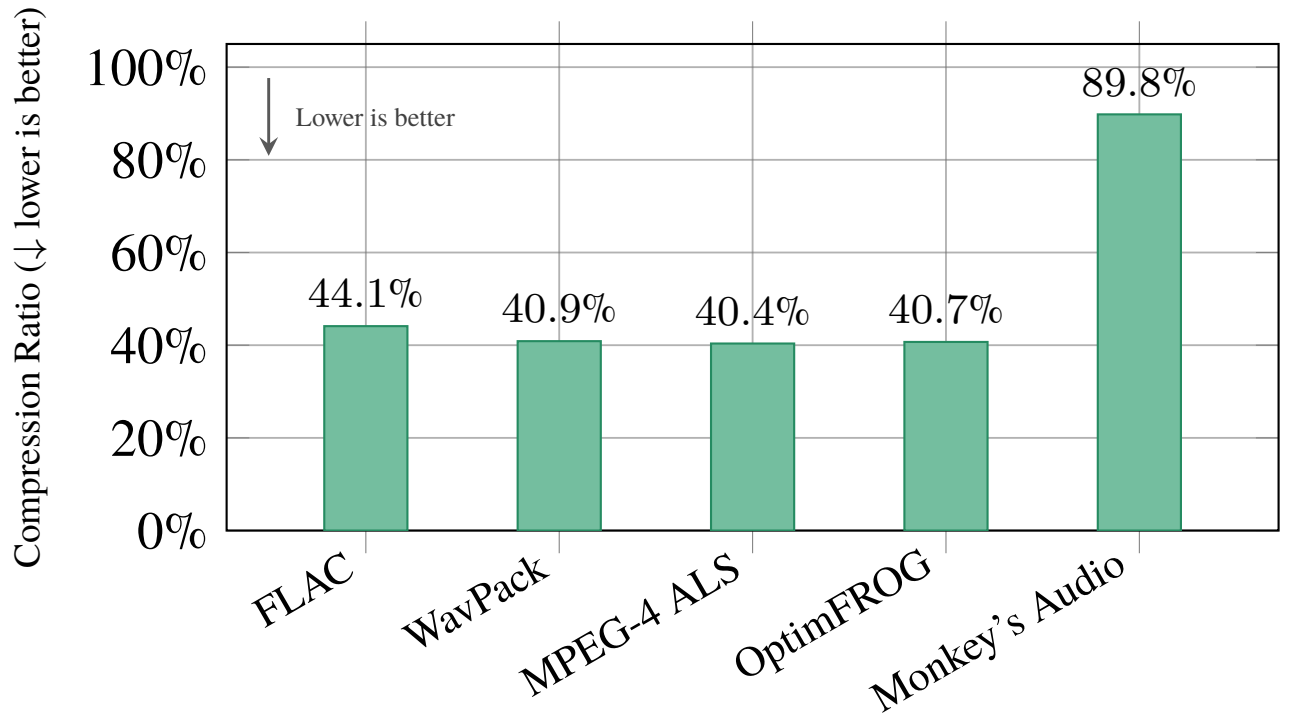

Because standard FLAC compresses 24-bit integer PCM to an average ratio of $\approx 60.2\%$, and DPA transforms the 32-bit floating-point stream into an aligned integer representation of equivalent effective precision, the final compression ratio evaluates to $\text{Ratio} \approx 60\% \times \frac{24}{32} = 45\%$. Thus, the storage footprint of raw 32-bit float audio is theoretically grounded to match that of conventional 24-bit integer PCM (see Fig.~\ref{fig:chart_adc}).

Additionally, engineers designing dual-stage converters are advised to calibrate analog preamplification to exact power-of-two voltage gain ratios ($A_v = 2^k$, such as $2\times, 4\times, 8\times$) rather than nominal decibel targets (such as $6$\,dB $\approx 1.995\times$, or arbitrary steps like $3$\,dB or $5$\,dB). Because decibels are base-10 logarithmic quantities, nominal decibel gain steps require fractional digital multipliers that scatter non-zero rounding noise across the entire mantissa. In contrast, an exact $2^k$ analog gain ratio allows digital firmware to stitch converter stages by merely adjusting the IEEE 754 exponent by $k$, leaving the significand bits bit-for-bit intact and preserving clean trailing zeros for optimal DPA packing.

\subsection{Workstation Masters and Consumer Delivery}
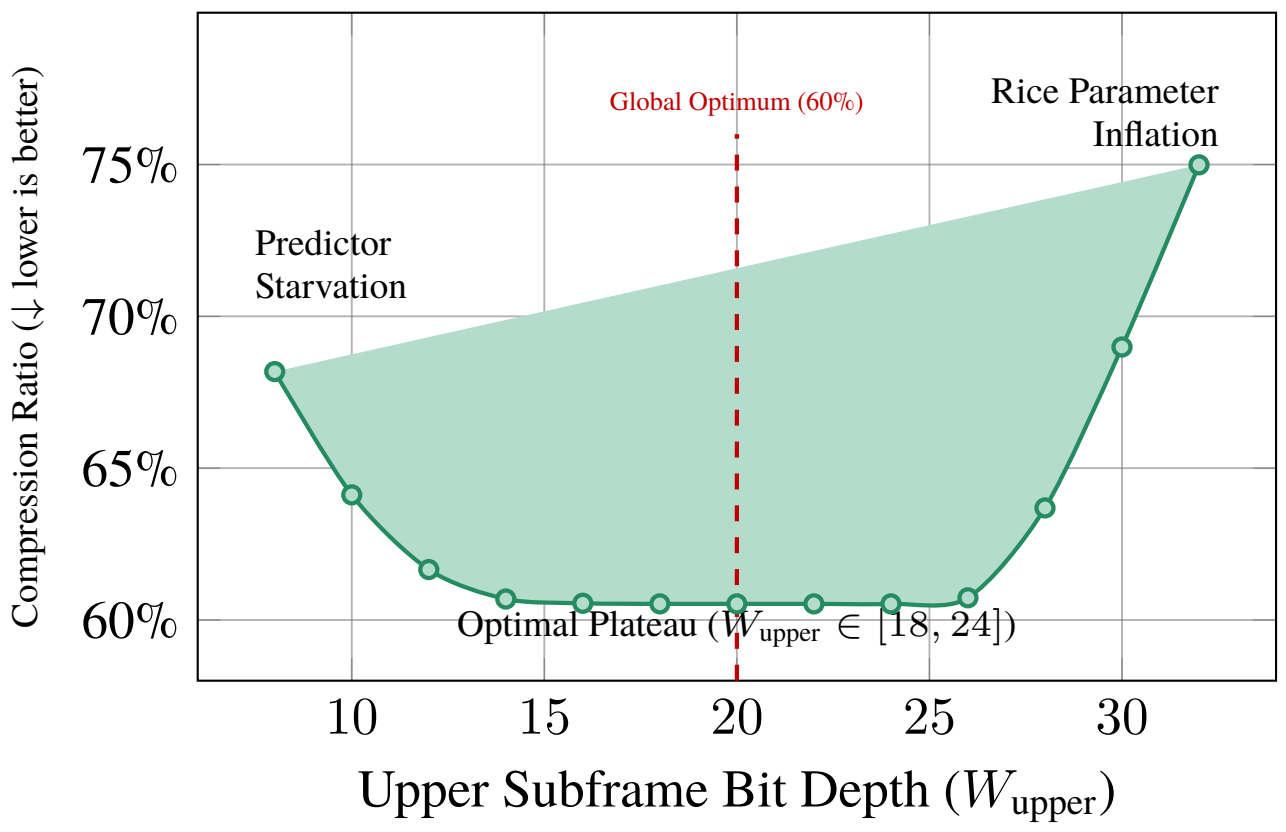
\begin{figure}[!ht]
	\centering
	\begin{tikzpicture}
		\begin{axis}[
				width=\columnwidth,
				height=6cm,
				xlabel={Upper Subframe Bit Depth ($W_{\text{upper}}$)},
				ylabel={Compression Ratio ($\downarrow$ lower is better)},
				ylabel style={font=\scriptsize},
				grid=both,
				grid style={line width=0.2pt, draw=black!40, opacity=0.45},
				major grid style={line width=0.35pt, draw=black!55, opacity=0.55},
				xmin=6, xmax=34,
				ymin=0.58, ymax=0.80,
				ytick={0.60, 0.65, 0.70, 0.75},
				yticklabels={60\%, 65\%, 70\%, 75\%},
				mark options={scale=0.8}
			]

			\addplot[thick, draw=wavegreen!90!black, mark=*, fill=wavegreen!35, smooth] coordinates {
					(8, 0.6818) (10, 0.6412) (12, 0.6166) (14, 0.6069)
					(16, 0.6055) (18, 0.6053) (20, 0.6053) (22, 0.6053)
					(24, 0.6053) (26, 0.6073) (28, 0.6369) (30, 0.6899) (32, 0.7499)
				};

			\node[anchor=south west, font=\scriptsize, align=left] at (axis cs:7, 0.70) {Predictor\\Starvation};
			\node[anchor=south, font=\scriptsize] at (axis cs:20, 0.585) {Optimal Plateau ($W_{\text{upper}} \in [18, 24]$)};
			\node[anchor=south east, font=\scriptsize, align=right] at (axis cs:33, 0.75) {Rice Parameter\\Inflation};

			\draw[red!75!black, dashed, line width=0.8pt] (axis cs:20, 0.58) -- (axis cs:20, 0.76);
			\node[anchor=south, text=red!75!black, font=\tiny] at (axis cs:20, 0.76) {Global Optimum (60\%)};

		\end{axis}
	\end{tikzpicture}%
	\caption{Compression ratio as a function of the upper subframe bit depth ($W_{\text{upper}}$) on a complex DAW master (lower is better). Isolating the Laplacian acoustic data from the Uniform float noise yields optimal compression, avoiding both predictor starvation and Rice coder inflation.}
	\label{fig:sp_ucurve}
\end{figure}

On complex digital masters, the operational efficiency of DPA mode hinges directly on the selected upper subframe bit depth ($W_{\text{upper}}$). Fig.~\ref{fig:sp_ucurve} illustrates the compression ratio as a function of $W_{\text{upper}}$ on complex 32-bit floating-point master files, revealing a distinct U-curve governed by the tension between predictor precision and entropy coder inflation:

\begin{itemize}
	\item \textbf{Left Branch ($W_{\text{upper}} < 18$): Predictor Starvation.} The upper subframe is allocated insufficient bits to model the acoustic waveform smoothly. The unmodeled dynamics overflow into the lower subframe where they are stored uncompressed, drastically inflating file size.
	\item \textbf{Right Branch ($W_{\text{upper}} > 24$): Rice Parameter Inflation.} As $W_{\text{upper}}$ exceeds the physical 24-bit single-precision mantissa span, trailing float noise is pulled into the upper subframe. The resulting variance spikes force the Rice coder to raise $k$ across entire partitions, exponentially degrading compression.
	\item \textbf{The Sweet Spot ($W_{\text{upper}} \in [18, 24]$):} By setting the upper resolution near the true physical noise floor, the codec isolates the Laplacian data from the Uniform data, achieving a global optimum compression ratio of $\approx 60\%$ at $W_{\text{upper}}=20$.
\end{itemize}

The U-curve essentially demonstrates that the effective number of bits (ENOB) per sample is around 16. In typical audio workflows, background noise limit coherent acoustic resolution to this range; allocating predictor bits beyond this boundary simply captures the noise.

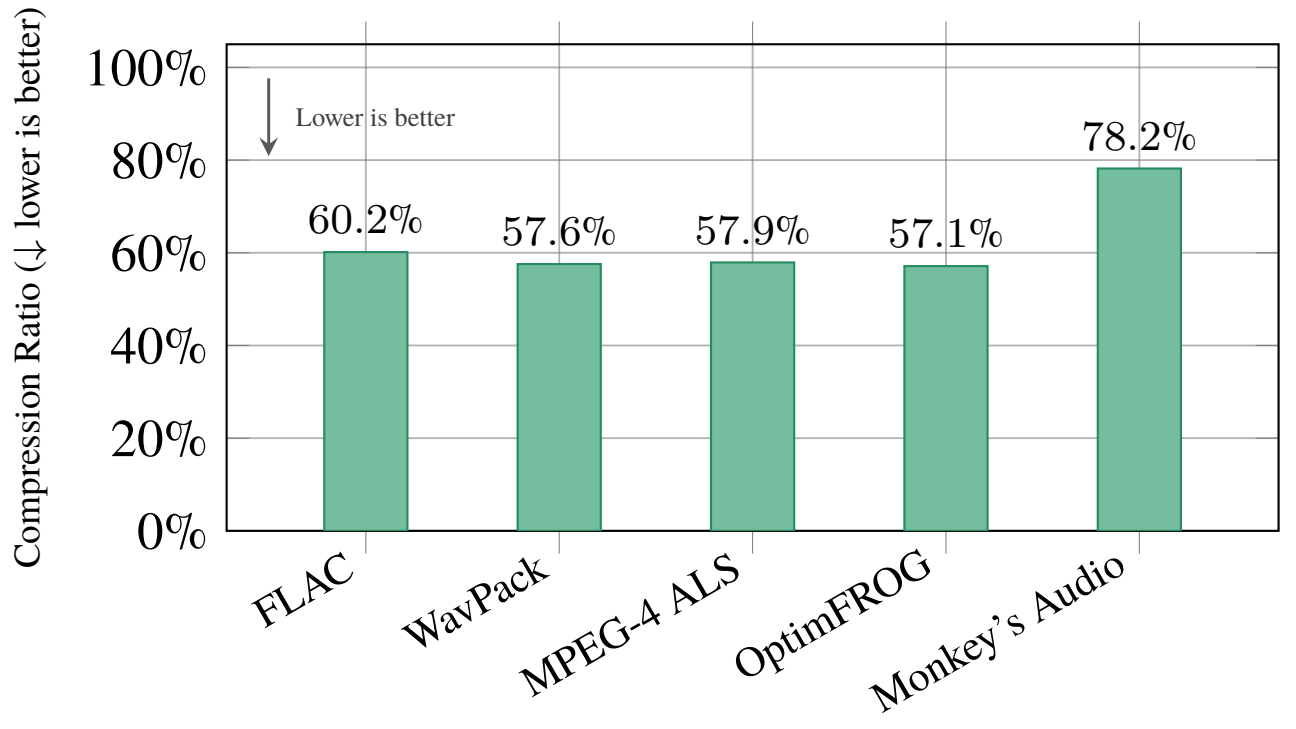
\begin{figure}[!ht]
	\centering
	\begin{tikzpicture}
		\begin{axis}[
				ybar,
				width=0.98\columnwidth,
				height=4.8cm,
				bar width=0.55cm,
				ymin=0, ymax=105,
				ytick={0, 20, 40, 60, 80, 100},
				yticklabels={0\%, 20\%, 40\%, 60\%, 80\%, 100\%},
				ylabel={Compression Ratio ($\downarrow$ lower is better)},
				ylabel style={font=\scriptsize},
				symbolic x coords={FLAC, WavPack, MPEG-4 ALS, OptimFROG, Monkey's Audio},
				xtick=data,
				nodes near coords={\pgfmathprintnumber\pgfplotspointmeta\%},
				nodes near coords style={font=\footnotesize, /pgf/number format/fixed, /pgf/number format/precision=1},
				grid=both,
				grid style={line width=0.2pt, draw=black!40, opacity=0.45},
				major grid style={line width=0.35pt, draw=black!55, opacity=0.55},
				enlarge x limits=0.18,
				x tick label style={font=\footnotesize, rotate=30, anchor=east}
			]
			\draw[->, >=stealth, line width=0.55pt, draw=black!65] (rel axis cs:0.04, 0.93) -- (rel axis cs:0.04, 0.77)
			node[midway, right=1.5pt, font=\tiny, text=black!75] {Lower is better};
			\addplot[fill=wavegreen!65, draw=wavegreen!90!black] coordinates {
					(FLAC, 60.15)
					(WavPack, 57.57)
					(MPEG-4 ALS, 57.92)
					(OptimFROG, 57.13)
					(Monkey's Audio, 78.19)
				};
		\end{axis}
	\end{tikzpicture}%
	\caption{Compression ratio across codecs under Complex DAW simulation on high-resolution musical material (Tier~3, 24-bit, $>48$\,kHz).}
	\label{fig:chart_daw}
\end{figure}

Theoretically, for uniformly distributed floating-point mantissas across full scale, half of the samples have their most significant bit active (requiring 1 packed bit), one quarter require 2 packed bits, and so on. The expected bit allocation across a 24-bit mantissa evaluates to:
\begin{equation}
	\sum_{i=1}^{24} i \cdot 2^{-i} \approx 2
\end{equation}
Because this geometric series converges strictly to~$2$, the asymptotic theoretical expectation is $2$\,bits per sample in the packed subframe. In real-world audio, empirical measurements yield an average of $\approx 3.6$\,bps. Because music is inherently non-stationary---exhibiting continuous dynamic fluctuations between quiet passages and loud transients---the active bit span varies across frames. Nevertheless, this bounded overhead confirms that raw serialization of fractional noise incurs low storage penalty. In particular, standard FLAC compresses the upper 24-bit audio subframe to an average ratio of $\approx 60\%$ ($14.4$\,effective bps). Adding the $\approx 3.6$\,bps lower subframe yields a combined bit budget of $18 / 32 \approx 56\%$. In practice, empirical evaluation on complex master material converges to an average compression ratio of $\approx 60.2\%$ (see Fig.~\ref{fig:chart_daw}).

For consumer delivery, FLAC introduces the dedicated export option:
\begin{center}
	\vspace{-0.5ex}
	\footnotesize\texttt{--force-empty-packed-subframe-which-is-not-lossless}
	\vspace{-0.5ex}
\end{center}
Applying this flag omits the \texttt{PACKED} subframe entirely, purging the uniform floating-point arithmetic noise floor while discarding virtually zero ENOB or real acoustic information.

Crucially, using this option produces an output stream whose file size is indistinguishable from a conventional 24-bit integer PCM stream. Online high-resolution music storefronts can therefore deliver native floating-point audio with zero storage overhead compared to their existing fixed-point catalogs. This provides three foundational industry benefits:
\begin{enumerate}
	\item \textbf{Clipping is Not Permanently Baked In:} Mastering engineers are freed from the artificial constraint of the 0\,dBFS brickwall limiter ceiling, preserving natural transient punch for downstream attenuation.
	\item \textbf{Superior Quality for Float ADC Recordings:} Acoustic captures from 32-bit float location recorders retain uncompressed dynamic fidelity without truncation or requantization artifacts.
	\item \textbf{End-to-End Floating-Point Pipeline:} Maintaining an unbroken floating-point workflow from tracking through mixing, mastering, and final delivery eliminates lossy format conversions and cumulative quantization errors.
\end{enumerate}

\subsection{Scientific Time Series and High-Precision Measurements}
\begin{figure}[!ht]
	\centering
	\begin{tikzpicture}
		\begin{axis}[
				ybar,
				width=0.98\columnwidth,
				height=4.8cm,
				bar width=0.55cm,
				ymin=0, ymax=140,
				ytick={0, 20, 40, 60, 80, 100, 120},
				yticklabels={0\%, 20\%, 40\%, 60\%, 80\%, 100\%, 120\%},
				ylabel={Compression Ratio ($\downarrow$ lower is better)},
				ylabel style={font=\scriptsize},
				symbolic x coords={FLAC, WavPack, MPEG-4 ALS, OptimFROG, Monkey's Audio},
				xtick=data,
				nodes near coords={\pgfmathprintnumber\pgfplotspointmeta\%},
				nodes near coords style={font=\footnotesize, /pgf/number format/fixed, /pgf/number format/precision=1},
				grid=both,
				grid style={line width=0.2pt, draw=black!40, opacity=0.45},
				major grid style={line width=0.35pt, draw=black!55, opacity=0.55},
				enlarge x limits=0.18,
				x tick label style={font=\footnotesize, rotate=30, anchor=east}
			]
			\draw[->, >=stealth, line width=0.55pt, draw=black!65] (rel axis cs:0.04, 0.93) -- (rel axis cs:0.04, 0.77)
			node[midway, right=1.5pt, font=\tiny, text=black!75] {Lower is better};
			\addplot[fill=wavegreen!65, draw=wavegreen!90!black] coordinates {
					(FLAC, 69.26)
					(WavPack, 72.44)
					(MPEG-4 ALS, 62.72)
					(OptimFROG, 67.61)
					(Monkey's Audio, 114.33)
				};
		\end{axis}
	\end{tikzpicture}%
	\caption{Macro-mean compression ratio across codecs under scientific time series and pathological signals (Tier~3; lower is better).}
	\label{fig:chart_scientific}
\end{figure}
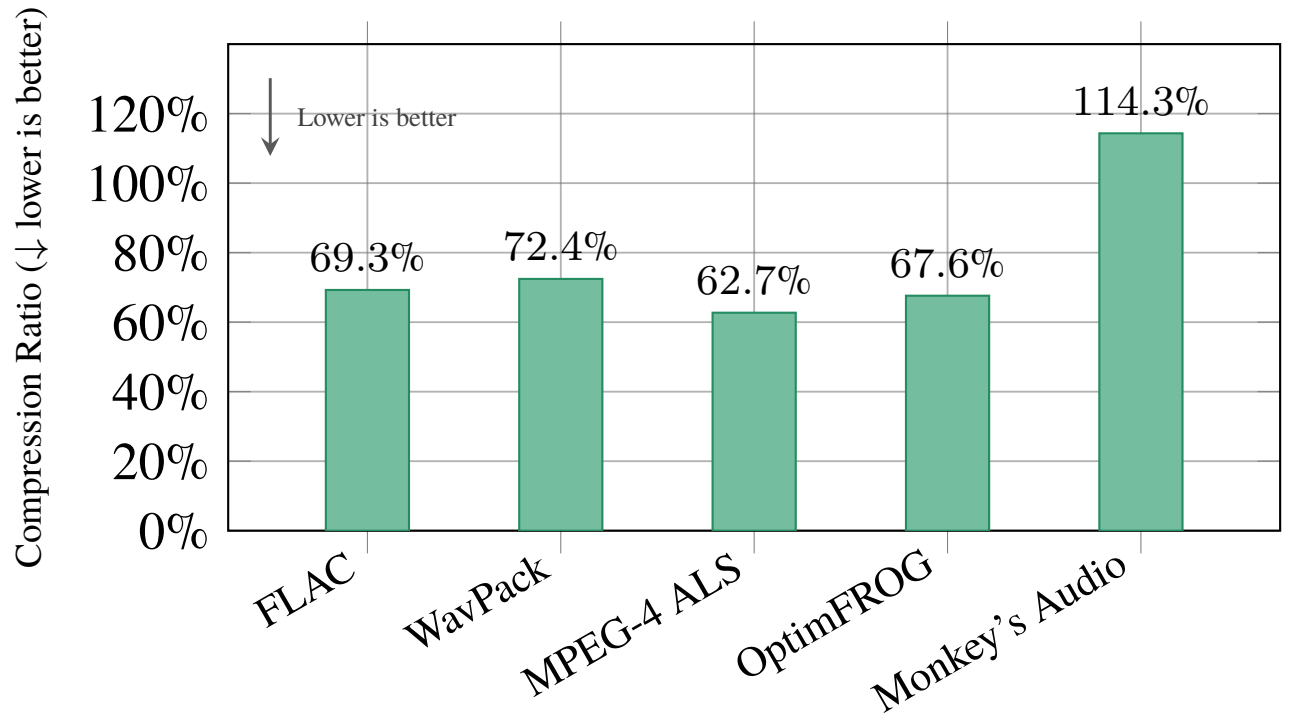

A $\approx 75\%$ compression ratio acts as a structural upper bound when FLAC adaptively reverts to fallback mode (see Fig.~\ref{fig:chart_scientific}). Because physical amplitude envelopes typically evolve smoothly over continuous trajectories, the 8-bit exponent subframe exhibits strong sample-to-sample autocorrelation and compresses to near-zero bitrate under linear prediction. Consequently, the compressed frame size is bounded almost entirely by the 24-bit significand subframe ($24/32 = 75\%$), even when sensor noise, steep decay, or high-entropy measurement dynamics render fractional bits incompressible.

Adaptive fallback also serves as an automated safeguard against pathological dynamic spans that defeat positional alignment. When intra-frame dynamics exceed the $31$-bit DPA window ($W_{\text{lower}} \le 31$), pure alignment would force out-of-window samples into uncompressed $32$-bit raw float serialization, expanding the payload beyond raw PCM. Instead, the encoder's efficiency threshold ($>75\%$ uncompressed payload) automatically reverts those frames to Fallback mode, maintaining bounded file sizes across all arbitrary inputs.

This architectural resilience allows libFLAC to reliably bound worst-case storage overhead on extreme signals, outperforming competitors overall without specialized format branches. A detailed breakdown across the synthetic stress regimes, along with an evaluation of non-finite sentinel telemetry, is provided in Appendix~\ref{sec:scientific_benchmarks}.

\section{Architectural Evolution and Design Trade-Offs}

Developing native lossless floating-point compression within libFLAC required navigating the trade-off between architectural complexity and compression ratio. To isolate the exact mathematical bottlenecks of IEEE 754 encoding, we evaluated five developmental stages spanning six distinct techniques before arriving at Dynamic Positional Alignment (DPA) with 2-way splitting. We document this architectural progression to provide future codec designers with an empirical cost-benefit reference.

\begin{enumerate}
	\item \textbf{Raw Floats as 32-Bit Signed Integers (Baseline):} The simplest conceivable approach feeds raw IEEE 754 binary32 bit patterns directly into FLAC's existing integer linear predictive coding (LPC) and Rice coding pipeline. Despite sign-magnitude discontinuities and step changes across power-of-two boundaries, standard integer predictors achieve an unoptimized compression ratio of $\approx 89.3\%$. This establishes the baseline performance floor requiring zero format modifications.

	\item \textbf{Single-Stream Bitwise Transformations:} To improve sample-to-sample autocorrelation without splitting channels into auxiliary subframes, we evaluated in-place bitwise mappings that transform binary32 words into pseudo-two's-complement integers prior to LPC analysis:
	      \begin{itemize}
		      \item \textit{Monkey's Audio (MAC) Transform:} In standard normalized audio, sample magnitudes strictly reside within $(-1, 1)$, meaning biased exponents are always $< 127$ (the highest bit is 0). Typical exponents cluster between 123 and 126, meaning the next 3 to 4 highest bits are consistently set to 1. The MAC \cite{MonkeysAudio} transform leverages this by applying the \texttt{0x3C000000} bitmask via XOR, which negates these predictable ``always 1'' bits. This effectively maps large exponent values down to a tight integer range (e.g., 0 to 7) at the top of the word. Coupled with conditional negation on the sign bit, this heuristic smooths the top bits for the integer predictor.
		      \item \textit{Cyclic Shift Transform:} While actual bit-shifting perfectly aligns mantissas, doing so within a fixed 32-bit word is inherently lossy. This transform attempted a lossless approximation via modular arithmetic. By cyclically shifting the significand by a modulo of the exponent (\nolinkurl{(23*12 - exponent)%23}), mantissas are dynamically aligned within any 23-wide exponent interval, causing lower exponents to align with themselves. While highly effective on clean 24-bit streams, it failed on complex DAW masters: cyclic shifting wrapped the incompressible floating-point rounding noise from the bottom of the mantissa directly into the high-order bits, destroying LPC correlation and performing worse than the raw float baseline.
		      \item \textit{Pseudo-Two's Complement Transform:} The optimal single-stream heuristic avoids cyclic bit-wrapping. It inverts the biased exponent ($e_{\text{inv}}=255-e$), maps it to the upper 8 bits, and conditionally complements the significand when the sign is negative. Because the exponent bits are mapped to the most significant bit positions, their smooth, slow-moving trajectories are almost guaranteed to be compressed highly efficiently by standard predictive coding. Furthermore, conditional negation enforces monotonic continuity across zero crossings. This achieves the highest compression ratio among single-stream techniques. However, transitions across power-of-two amplitude boundaries remain the central bottleneck for the coder.
	      \end{itemize}

	\item \textbf{Forced 2-Subframe Fallback Mode:} Decouples each channel into an 8-bit Exponent subframe (centered with a $-128$ DC offset) and a 24-bit Significand subframe mapped to signed two's complement notation. This removes step discontinuities and establishes a continuous exponent trajectory that facilitates inter-channel decorrelation and LPC filtering for the exponent bits. However, lacking global positional alignment, the LPC filter must model unaligned mantissa amplitudes independently, yielding overall poor efficiency.

	\item \textbf{Dynamic Positional Alignment (Base Alignment in 1 Subframe Only):} To allow standard predictors to operate on true time-domain values, this iteration aligns all frame samples to the lowest active position ($P_{\text{low}}$) via left-shifting. On multi-stage ADC recordings, the active bit span remains strictly bounded by 24 bits ($W_{\text{eff}} \le 24$), allowing the standard LPC pipeline to compress the stream to $\approx 43.1\%$ with zero structural modifications. On complex DAW masters, active bit spans expand up to 47 bits. To align to the quietest sample, loud samples (with higher exponents) must be left-shifted significantly. While the predictor perfectly models the scaled waveform, this left-shift physically pads the loud samples with large quantities of uninformative trailing zeros (e.g., up to 23 zeros). The entropy coder is forced to physically serialize these trailing zeros, severely bottlenecking compression.

	\item \textbf{DPA with Shift-Normalized Prediction (Delta Stream):} To eliminate the entropy overhead of storing the trailing zeros generated by loud samples, this iteration introduced a dedicated 8-bit Delta subframe ($\Delta_i$) to serialize exponent shift trajectories, enabling shift-normalized prediction. Aligning samples via a left shift ($x_i=(-1)^{s_i}\cdot(M_i\ll\Delta_i)$) establishes a strict mathematical congruence constraint on the integer signal. This relationship proves that the bottom $\Delta_i$ bits of the prediction residual are completely deterministic given the predictor state $\hat{x}_i$. Rather than spending entropy-coding bandwidth transmitting these redundant low-order bits, the encoder extracts the deterministic remainder $R_i=(-\hat{x}_i)\&(2^{\Delta_i}-1)$ and normalizes the residual via an arithmetic right shift $Q_i=r_i\gg\Delta_i$. At decode time, the decoder decodes the Delta subframe to recover $\Delta_i$, computes $\hat{x}_i$ from reconstructed prior samples, evaluates $R_i$, and restores the exact residual $r_i=(Q_i\ll\Delta_i)|R_i$.

	      While shift normalization successfully stripped the trailing zeros from the loud samples, empirical instrumentation revealed that the quotient residuals $Q_i$ formed a rigid, uniform floor of $\approx 17\text{ bits/sample}$ across entire blocks. Normalizing residuals by dividing by the local sample shift mathematically scaled the prediction errors back to the full 24-bit mantissa domain. Because non-power-of-two gain staging fills the lower mantissa of every sample with uniform rounding noise, feeding this uniform distribution into a partitioned Rice coder caused partition-wide parameter inflation, capping DAW compression at $\approx 71.3\%$.

	\item \textbf{DPA with 2-Way Splitting (Current Work):} Solves the uniform noise bottleneck through pre-prediction distribution segregation. Rather than normalizing residuals post-prediction, DPA segregates the aligned integer into an Upper Subframe of width $W_{\text{upper}}$ ($x_{\text{upper}}=x_i\gg W_{\text{lower}}$) and a Lower Subframe ($x_{\text{lower}}=x_i\ \&\ ((1\ll W_{\text{lower}})-1)$), with lower bit depth bounded by $W_{\text{lower}}\le 31$. Outliers and samples with $x_{\text{upper}}=0$ (including zeros, subnormals, and non-finite floats) are directly captured as raw 32-bit floats within the auxiliary stream. The Upper Subframe preserves the Laplacian acoustic distribution for optimal LPC and Rice coding, while the Lower Subframe serializes uniform rounding noise as uncompressed packed bits, completely bypassing Rice coding.

	\item \textbf{Common Factor Estimation and Dictionary Coding (MPEG-4 ALS):} Beyond bitwise transformations and DPA, MPEG-4 ALS (ISO/IEC 14496-3) \cite{mp4} adopts two fundamentally different mechanisms for floating-point audio: Common Factor Estimation (CFE) and Lempel-Ziv (LZ) compression of lower mantissa bits. CFE operates under the premise that the audio stream originates from a 24-bit integer source with an added uniform gain factor that alters the bit structure; it attempts to identify this common multiplier and invert it to recover discrete integer samples. However, this assumption is valid only for simple gain scaling. Production DAWs are substantially more complex: a single equalization (EQ) filter, dynamic compressor, or multi-bus summing pass thoroughly breaks this assumption, rendering CFE useless. Similarly, LZ dictionary compression is ineffective on complex workstation masters. LZ compression only yields meaningful gains when simplistic arithmetic operations have been applied to the audio stream (e.g., dividing by 3, which generates repeating binary bit sequences across samples). Under such contrived conditions, our internal results suggested that MPEG-4 ALS achieves compression ratios near $\approx 45\%$ while other techniques remain near $\approx 60\%$. However, across authentic real-world masters, fractional arithmetic noise is statistically uniform and non-repetitive, offering no recurrent substrings for dictionary parsing. Our benchmarks on realistic production material confirm that these techniques provide no practical advantage on real-world streams. Consequently, adding this algorithmic complexity to the FLAC codec specification was not considered.
\end{enumerate}

\begin{figure}[!ht]
	\centering
	\begin{tikzpicture}
		\begin{axis}[
				ybar,
				width=0.98\columnwidth,
				height=5.6cm,
				bar width=0.15cm,
				ymin=0, ymax=115,
				ytick={0, 20, 40, 60, 80, 100},
				yticklabels={0\%, 20\%, 40\%, 60\%, 80\%, 100\%},
				ylabel={Compression Ratio ($\downarrow$ lower is better)},
				ylabel style={font=\scriptsize},
				symbolic x coords={Raw, CyclicShift, BitManip, Fallback, OneSubframe, DPA-SNP, DPA},
				xtick=data,
				nodes near coords={\pgfmathprintnumber\pgfplotspointmeta\%},
				nodes near coords style={font=\tiny, rotate=90, anchor=west, /pgf/number format/fixed, /pgf/number format/precision=1},
				grid=both,
				grid style={line width=0.2pt, draw=black!40, opacity=0.45},
				major grid style={line width=0.35pt, draw=black!55, opacity=0.55},
				enlarge x limits=0.10,
				x tick label style={font=\scriptsize, rotate=40, anchor=east},
				legend style={
						at={(0.5, 1.03)},
						anchor=south,
						legend columns=2,
						font=\scriptsize,
						draw=none,
						/tikz/every even column/.append style={column sep=0.4cm}
					}
			]
			\draw[->, >=stealth, line width=0.55pt, draw=black!65] (rel axis cs:0.96, 0.94) -- (rel axis cs:0.96, 0.78)
			node[midway, left=1.5pt, font=\tiny, text=black!75] {Lower is better};
			\addplot[fill=wavegreen!65, draw=wavegreen!90!black] coordinates {
					(Raw, 70.01)
					(CyclicShift, 78.81)
					(BitManip, 73.05)
					(Fallback, 68.23)
					(OneSubframe, 43.10)
					(DPA-SNP, 51.00)
					(DPA, 44.11)
				};
			\addlegendentry{Dual-Stage ADC}

			\addplot[fill=svgmaroon!40, draw=svgmaroon!90!black] coordinates {
					(Raw, 89.33)
					(CyclicShift, 88.46)
					(BitManip, 71.58)
					(Fallback, 74.84)
					(OneSubframe, 74.94)
					(DPA-SNP, 71.27)
					(DPA, 60.15)
				};
			\addlegendentry{Complex DAW Master}
		\end{axis}
	\end{tikzpicture}%
	\caption{Compression ratios across architectural evolution stages under Dual-Stage ADC and Complex DAW simulations on high-resolution musical material (Tier~3, 24-bit, $>48$\,kHz).}
	\label{fig:chart_evolution}
\end{figure}
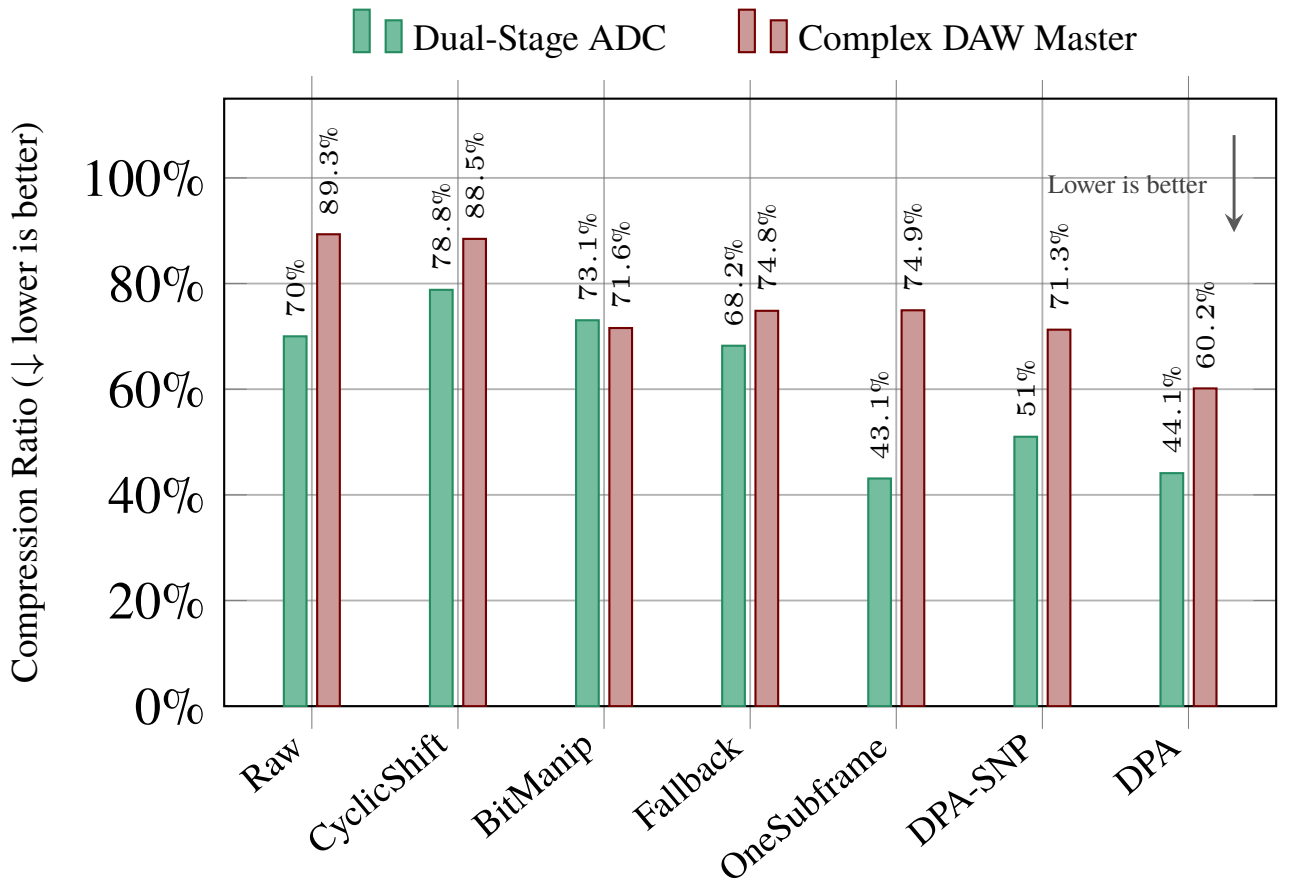

\section*{Acknowledgments}
The author would like to express sincere gratitude to Prof.~Mohammad Hossein Rohban\footnote{Department of Computer Engineering, Sharif University of Technology. Email: \href{mailto:rohban@sharif.edu}{\texttt{rohban@sharif.edu}}.} and Prof.~Mohammad Bahrani\footnote{Faculty of Mathematical Sciences and Computer, Allameh Tabataba\'i University. Email: \href{mailto:bahrani@atu.ac.ir}{\texttt{bahrani@atu.ac.ir}}.} for their supervision, guidance, and support throughout this research.

\printbibliography

\appendix

\section{Experimental Evaluation and Benchmarks}
\label{sec:benchmarks}

To evaluate the empirical efficiency of the proposed floating-point implementation, we benchmarked our native FLAC fork against the established state of the art in lossless floating-point audio compression: \textbf{WavPack 5.9.0}, \textbf{MPEG-4 ALS v23}, \textbf{OptimFROG (off) v5.1000}, and \textbf{Monkey's Audio 13.26}.

\subsection{Benchmark Corpus and Simulation Models}
The evaluation suite comprises a three-tier corpus totaling 144 audio tracks:
\begin{itemize}
	\item \textbf{Tier 1: Canonical Public Sets (101 tracks):} The complete official European Broadcasting Union Sound Quality Assessment Material (EBU SQAM, Tech 3253; 70 tracks, 16-bit / 44.1\,kHz) and the complete public-domain studio recording of J.S. Bach's \textit{Open Goldberg Variations} performed by Kimiko Ishizaka (31 tracks, native 24-bit studio masters).
	\item \textbf{Tier 2: Stratified Real-World Music (40 tracks):} Commercial master recordings sampled across ten diverse genres (including blues, rock, electronic, jazz, progressive metal, soul, and acoustic pop).
	\item \textbf{Tier 3: Scientific Time Series and Pathological Signals (3 tracks):} Algorithmic and sensor-telemetry stress models (including wide-range exponent staircases, exponential seismic decay, and sentinel NaN dropouts).
\end{itemize}

All audio tracks across the three tiers were originally fixed-point integer PCM streams (16-bit or 24-bit). Because native floating-point distribution files are not yet commercially widespread, we converted these integer streams to IEEE 754 single-precision float and constructed two deterministic operational simulations:
\begin{enumerate}
	\item \textbf{Dual-Stage ADC Simulation:} To realistically simulate a 2-stage ADC architecture, the audio stream is subjected to discrete gain staging. Every 10 seconds, samples are shifted by an exact power-of-two factor of $2^5$ ($32\times$) only, with a 50\% probability. Because exact power-of-two scaling modifies only the IEEE 754 exponent field, the significand bits remain bit-for-bit unaltered outside stage transitions.
	\item \textbf{Complex DAW Master Simulation:} Each float sample $x \in [-1.0, 1.0]$ is transformed via a non-linear power curve $x \mapsto \operatorname{sgn}(x) \cdot \vert{}x\vert{}^{0.9}$. This non-linearity simulates complex digital workstation processing (such as equalization and dynamic shaping) that thoroughly alters the floating-point mantissa bits while keeping the samples relatively bounded within the normalized $[-1.0, 1.0]$ interval. Under this simulation, arithmetic rounding thoroughly populates all 23 significand bits, expanding active bit spans to 39 bits for 16-bit sources and 47 bits for 24-bit sources, triggering dynamic two-way splitting.
\end{enumerate}

All codecs were verified to decompress $100\%$ bit-for-bit losslessly across all 144 tracks and both simulation regimes.

\subsection{Operational Throughput (Real-Time Factors)}
To evaluate operational throughput realistically for interactive audio workstations and consumer battery-constrained devices, speed comparisons were conducted using standard (default) compression presets across all codecs: \texttt{flac -5}, \texttt{wavpack} (default), \texttt{mp4als} (default), \texttt{off} (default), and \texttt{mac -c2000}. Throughput was measured in real-time playback factors ($\times\text{ real-time}$, defined as stream duration divided by processing duration).

\begin{table}[!ht]
	\centering
	\caption{Operational Throughput Comparison at Default Presets}
	\label{tab:throughput_comparison}
	\resizebox{\columnwidth}{!}{%
		\begin{tabular}{@{}llcc@{}}
			\toprule
			\textbf{Simulation Scenario} & \textbf{Codec}                    & \textbf{Encode Speed}   & \textbf{Decode Speed}   \\
			                             &                                   & ($\times$ Real-Time)    & ($\times$ Real-Time)    \\ \midrule
			\multirow{5}{*}{Dual-Stage ADC}
			                             & \textbf{FLAC (Ours, \texttt{-5})} & \textbf{169.1}$\times$  & \textbf{160.9}$\times$  \\
			                             & WavPack (Default)                 & \phantom{0}$62.4\times$ & $117.9\times$           \\
			                             & MPEG-4 ALS (Default)              & \phantom{0}$20.1\times$ & \phantom{0}$66.0\times$ \\
			                             & OptimFROG (Default)               & \phantom{0}$18.5\times$ & \phantom{0}$27.5\times$ \\
			                             & Monkey's Audio (\texttt{-c2000})  & \phantom{0}$33.2\times$ & \phantom{0}$31.9\times$ \\ \cmidrule(lr){1-4}
			\multirow{5}{*}{Complex DAW Master}
			                             & \textbf{FLAC (Ours, \texttt{-5})} & \textbf{114.6}$\times$  & $101.0\times$           \\
			                             & WavPack (Default)                 & \phantom{0}$53.9\times$ & \textbf{103.6}$\times$  \\
			                             & MPEG-4 ALS (Default)              & \phantom{00}$5.8\times$ & \phantom{0}$57.8\times$ \\
			                             & OptimFROG (Default)               & \phantom{0}$17.2\times$ & \phantom{0}$24.6\times$ \\
			                             & Monkey's Audio (\texttt{-c2000})  & \phantom{0}$34.7\times$ & \phantom{0}$34.1\times$ \\ \bottomrule
		\end{tabular}%
	}
\end{table}

\subsection{Cross-Codec Compression Benchmarks}
Tables~\ref{tab:compression_ratio_adc} and~\ref{tab:compression_ratio_daw} present compression ratios across the corpus. For evaluating compression ratios, all codecs were evaluated at their highest compression settings: \texttt{flac -8}, \texttt{wavpack -hh -x6} \texttt{--threads=1}, \texttt{mp4als -7 -f5}, \texttt{off --mode} \texttt{highnew\allowbreak -light} \texttt{--optimize best}, and \texttt{mac -c5000 -threads=1}.

\begin{table*}[!ht]
	\centering
	\caption{Lossless Compression Ratios for Dual-Stage ADC Simulation (Maximum Presets; lower is better)}
	\label{tab:compression_ratio_adc}
	\resizebox{\textwidth}{!}{%
		\begin{tabular}{@{}lllcccccccc@{}}
			\toprule
			\textbf{Corpus Tier} & \textbf{Format Strata}                                     & \textbf{Sampling Rate} & \textbf{FLAC (Ours)} & \textbf{WavPack} & \textbf{MPEG-4 ALS} & \textbf{OptimFROG}     & \textbf{Monkey's Audio 13.26} & \multicolumn{3}{c}{\textbf{Paired Difference vs. WavPack}}                                                   \\ \cmidrule(l){9-11}
			                     &                                                            &                        & (Mean $\pm$ 90\% CI) & \texttt{-hh -x6} & \texttt{-7 -f5}     & \texttt{highnew-light} & \texttt{-c5000}               & $\mathbf{\Delta}$ (\%)                                     & \textbf{90\% CI}  & $\mathbf{p}$\textbf{-value} \\ \midrule
			\multirow{2}{*}{\textbf{Tier 1: Canonical}}
			                     & 16-bit PCM                                                 & $\le 48$\,kHz          & $24.18 \pm 1.08$\%   & 14.11\%          & 12.67\%             & \textbf{12.22\%}       & 75.70\%                       & $+10.07\%$                                                 & $[+9.35, +10.79]$ & $< 10^{-4}$                 \\ \cmidrule(lr){2-11}
			                     & 24-bit PCM                                                 & $> 48$\,kHz            & $35.52 \pm 0.25$\%   & 35.20\%          & \textbf{34.57\%}    & 35.28\%                & 87.51\%                       & $+0.32\%$                                                  & $[+0.31, +0.33]$  & $< 10^{-4}$                 \\ \midrule
			\multirow{4}{*}{\textbf{Tier 2: Stratified Music}}
			                     & 16-bit PCM                                                 & $\le 48$\,kHz          & $34.43 \pm 1.06$\%   & 30.78\%          & 30.43\%             & \textbf{29.78\%}       & 104.48\%                      & $+3.65\%$                                                  & $[+3.11, +4.20]$  & $< 10^{-4}$                 \\ \cmidrule(lr){2-11}
			                     & \multirow{2}{*}{24-bit PCM}                                & $\le 48$\,kHz          & $55.84 \pm 8.76$\%   & 54.01\%          & 53.14\%             & \textbf{53.08\%}       & 97.36\%                       & $+1.83\%$                                                  & $[+0.72, +2.94]$  & $0.030$                     \\ \cmidrule(lr){3-11}
			                     &                                                            & $> 48$\,kHz            & $44.11 \pm 2.93$\%   & 40.87\%          & \textbf{40.36\%}    & 40.70\%                & 89.81\%                       & $+3.23\%$                                                  & $[+0.63, +5.83]$  & $0.046$                     \\ \cmidrule(lr){2-11}
			                     & \multicolumn{2}{l}{\textit{Tier 2 Macro-Mean (40 Tracks)}} & $39.96 \pm 2.24$\%     & 36.63\%              & 36.17\%          & \textbf{35.93\%}    & 98.63\%                & $+3.32\%$                     & $[+2.42, +4.23]$                                           & $< 10^{-4}$                                     \\ \bottomrule
		\end{tabular}%
	}
\end{table*}

\begin{table*}[!ht]
	\centering
	\caption{Lossless Compression Ratios for Complex DAW Master Simulation (Maximum Presets; lower is better)}
	\label{tab:compression_ratio_daw}
	\resizebox{\textwidth}{!}{%
		\begin{tabular}{@{}lllcccccccc@{}}
			\toprule
			\textbf{Corpus Tier} & \textbf{Format Strata}                                     & \textbf{Sampling Rate} & \textbf{FLAC (Ours)} & \textbf{WavPack} & \textbf{MPEG-4 ALS} & \textbf{OptimFROG}     & \textbf{Monkey's Audio 13.26} & \multicolumn{3}{c}{\textbf{Paired Difference vs. WavPack}}                                                  \\ \cmidrule(l){9-11}
			                     &                                                            &                        & (Mean $\pm$ 90\% CI) & \texttt{-hh -x6} & \texttt{-7 -f5}     & \texttt{highnew-light} & \texttt{-c5000}               & $\mathbf{\Delta}$ (\%)                                     & \textbf{90\% CI} & $\mathbf{p}$\textbf{-value} \\ \midrule
			\multirow{2}{*}{\textbf{Tier 1: Canonical}}
			                     & 16-bit PCM                                                 & $\le 48$\,kHz          & $54.46 \pm 1.55$\%   & 51.01\%          & \textbf{50.07\%}    & 50.31\%                & 68.74\%                       & $+3.46\%$                                                  & $[+2.41, +4.50]$ & $< 10^{-4}$                 \\ \cmidrule(lr){2-11}
			                     & 24-bit PCM                                                 & $> 48$\,kHz            & $54.97 \pm 0.24$\%   & 54.47\%          & 54.77\%             & \textbf{54.16\%}       & 83.06\%                       & $+0.50\%$                                                  & $[+0.43, +0.57]$ & $< 10^{-4}$                 \\ \midrule
			\multirow{4}{*}{\textbf{Tier 2: Stratified Music}}
			                     & 16-bit PCM                                                 & $\le 48$\,kHz          & $70.71 \pm 1.23$\%   & 68.61\%          & 68.37\%             & \textbf{67.83\%}       & 81.80\%                       & $+2.10\%$                                                  & $[+1.67, +2.54]$ & $< 10^{-4}$                 \\ \cmidrule(lr){2-11}
			                     & \multirow{2}{*}{24-bit PCM}                                & $\le 48$\,kHz          & $69.65 \pm 5.67$\%   & 67.59\%          & 67.36\%             & \textbf{66.66\%}       & 80.62\%                       & $+2.06\%$                                                  & $[+0.33, +3.79]$ & $0.067$                     \\ \cmidrule(lr){3-11}
			                     &                                                            & $> 48$\,kHz            & $60.15 \pm 2.03$\%   & 57.57\%          & 57.92\%             & \textbf{57.13\%}       & 78.19\%                       & $+2.58\%$                                                  & $[-0.08, +5.23]$ & $0.109$                     \\ \cmidrule(lr){2-11}
			                     & \multicolumn{2}{l}{\textit{Tier 2 Macro-Mean (40 Tracks)}} & $66.91 \pm 1.67$\%     & 64.64\%              & 64.61\%          & \textbf{63.97\%}    & 80.42\%                & $+2.27\%$                     & $[+1.36, +3.17]$                                           & $< 10^{-4}$                                    \\ \bottomrule
		\end{tabular}%
	}
\end{table*}

\subsection[Scientific Time Series and Pathological Tests]{Scientific Time Series and\\Pathological Stress Tests}
\label{sec:scientific_benchmarks}

To evaluate codec robustness beyond acoustic material, we constructed Tier~3, consisting of synthetic stress models designed to evaluate extreme dynamic divergence, automated fallback switching, and non-finite representation. The resulting compression metrics are reported in Table~\ref{tab:scientific_benchmarks}.

\begin{table}[!ht]
	\centering
	\caption{Lossless Compression Ratios on Scientific Time Series and Pathological Signals (Tier~3; lower is better)}
	\label{tab:scientific_benchmarks}
	\resizebox{\columnwidth}{!}{%
		\begin{tabular}{@{}lccccc@{}}
			\toprule
			\textbf{Synthetic Signal} & \textbf{FLAC (Ours)} & \textbf{WavPack} & \textbf{MPEG-4 ALS} & \textbf{OptimFROG} & \textbf{Monkey's Audio} \\ \midrule
			Exponent Staircase        & 74.87\%              & 90.74\%          & \textbf{55.05\%}    & 82.82\%            & 71.94\%                 \\
			Seismic Ringdown          & \textbf{81.52\%}     & 93.59\%          & 90.31\%             & 83.16\%            & 87.53\%                 \\
			Sensor Dropout (NaNs)     & 77.92\%              & 52.24\%          & 48.87\%             & \textbf{44.07\%}   & 229.68\%                \\ \midrule
			\textbf{Macro-Mean}       & 78.10\%              & 78.86\%          & \textbf{64.74\%}    & 70.02\%            & 129.72\%                \\ \bottomrule
		\end{tabular}%
	}
\end{table}

\noindent\textbf{Evaluation Regimes and Architectural Rationale}

\begin{enumerate}
	\item \textbf{Monotonic Exponent Staircase:}
	      A single 4096-sample frame subdivided into 64 uniform blocks of 64 samples, stepping the unbiased exponent monotonically from 0 down to $-63$ (spanning 64 powers of two per frame).
	      \begin{itemize}
		      \item \textit{Rationale:} Deliberately exceeds the 31-bit positional register window ($W_{\text{lower}} \le 31$) within every single frame. This verifies that the encoder cleanly detects uncompressed literal bloat and reverts to Fallback mode. Because the piecewise-linear exponent staircase is easily modeled by LPC, the 8-bit exponent subframe collapses to near-zero bitrate, verifying the theoretical $24/32 \approx 75\%$ bound.
	      \end{itemize}

	\item \textbf{Seismic Shock and Exponential Decay:}
	      Simulates wide-range physical impulse response telemetry (such as seismological sensor events or nuclear magnetic resonance free induction decays), dropping 48 powers of two ($>280$\,dB) in amplitude within each 4096-sample frame.
	      \begin{itemize}
		      \item \textit{Rationale:} Evaluates adaptive switching during sharp decay transients where positional alignment would otherwise suffer severe literal overhead. The efficiency safeguard successfully confines the frame size below the raw PCM stream, outperforming both WavPack and Monkey's Audio.
	      \end{itemize}

	\item \textbf{Sensor Dropout Telemetry with Sentinel NaNs:}
	      Simulates instrumentation loss and missing sensor frames, introducing random 50\% dropouts where samples are independently replaced by canonical IEEE 754 quiet NaN sentinels (\texttt{0x7FC00000}).
	      \begin{itemize}
		      \item \textit{Rationale:} Verifies bit-exact preservation and stability on non-finite data. While WavPack achieves a smaller footprint by dedicating a single bit to flag a NaN and storing only its mantissa, FLAC stores the entire raw 32-bit float without exception, which triggers a fallback under high dropout rates. Although FLAC could adopt a scheme similar to WavPack for irregular numbers (such as NaNs and infinities), we deliberately decided not to. Avoiding special-case exceptions keeps the codec format specification simple, facilitating easier and faster adoption for third-party implementations beyond the reference codec.
	      \end{itemize}
\end{enumerate}

\subsection{Ablation Across Architectural Iterations}
\label{sec:ablation_benchmarks}

To evaluate the empirical impact of each design milestone detailed in Section~4, we systematically benchmarked all developmental iterations across every tier, format stratum, and sampling rate category in the evaluation corpus. Tables~\ref{tab:ablation_adc} and~\ref{tab:ablation_daw} document the resulting compression ratios under the Dual-Stage ADC and Complex DAW Master simulation models, respectively. These measurements trace the performance trajectory from the unoptimized raw float baseline through single-stream transforms and shift-normalized prediction, culminating in the current two-way split DPA architecture.

\begin{table*}[!ht]
	\centering
	\caption{Lossless Compression Ratios Across Architectural Variants for Dual-Stage ADC Simulation (\texttt{-8} Preset; lower is better)}
	\label{tab:ablation_adc}
	\resizebox{\textwidth}{!}{%
		\begin{tabular}{@{}lllcccccccc@{}}
			\toprule
			\textbf{Corpus Tier} & \textbf{Format Strata}                                     & \textbf{Sampling Rate} & \textbf{Raw} & \textbf{CyclicShift} & \textbf{BitManip} & \textbf{Fallback} & \textbf{OneSubframe} & \textbf{DPA\_SNP}  & \textbf{DPA (Ours)}  \\
			                     &                                                            &                        &              &                      &                   &                   &                      &                    & (Mean $\pm$ 90\% CI) \\ \midrule
			\multirow{2}{*}{\textbf{Tier 1: Canonical}}
			                     & 16-bit PCM                                                 & $\le 48$\,kHz          & 37.60\%      & 64.52\%              & 36.91\%           & 32.74\%           & \textbf{14.85\%}     & 21.21\%            & $24.18 \pm 1.08$\%   \\ \cmidrule(lr){2-10}
			                     & 24-bit PCM                                                 & $> 48$\,kHz            & 69.09\%      & 79.63\%              & 70.43\%           & 64.01\%           & \textbf{35.37\%}     & 43.32\%            & $35.52 \pm 0.25$\%   \\ \midrule
			\multirow{4}{*}{\textbf{Tier 2: Stratified Music}}
			                     & 16-bit PCM                                                 & $\le 48$\,kHz          & 58.48\%      & 85.39\%              & 59.90\%           & 49.30\%           & \textbf{32.30}\%     & 40.16\%            & $34.43 \pm 1.06$\%   \\ \cmidrule(lr){2-10}
			                     & \multirow{2}{*}{24-bit PCM}                                & $\le 48$\,kHz          & 79.93\%      & 84.71\%              & 81.78\%           & 72.75\%           & \textbf{55.62}\%     & 63.37\%            & $55.84 \pm 8.76$\%   \\ \cmidrule(lr){3-10}
			                     &                                                            & $> 48$\,kHz            & 70.01\%      & 78.81\%              & 73.05\%           & 68.23\%           & \textbf{43.10}\%     & 51.00\%            & $44.11 \pm 2.93$\%   \\ \cmidrule(lr){2-10}
			                     & \multicolumn{2}{l}{\textit{Tier 2 Macro-Mean (40 Tracks)}} & 64.66\%                & 83.02\%      & 66.69\%              & 58.27\%           & \textbf{38.42}\%  & 46.27\%              & $39.96 \pm 2.24$\%                        \\ \bottomrule
		\end{tabular}%
	}
\end{table*}

\begin{table*}[!ht]
	\centering
	\caption{Lossless Compression Ratios Across Architectural Variants for Complex DAW Master Simulation (\texttt{-8} Preset; lower is better)}
	\label{tab:ablation_daw}
	\resizebox{\textwidth}{!}{%
		\begin{tabular}{@{}lllcccccccc@{}}
			\toprule
			\textbf{Corpus Tier} & \textbf{Format Strata}                                     & \textbf{Sampling Rate} & \textbf{Raw} & \textbf{CyclicShift} & \textbf{BitManip} & \textbf{Fallback} & \textbf{OneSubframe} & \textbf{DPA\_SNP}           & \textbf{DPA (Ours)}         \\
			                     &                                                            &                        &              &                      &                   &                   &                      &                             & (Mean $\pm$ 90\% CI)        \\ \midrule
			\multirow{2}{*}{\textbf{Tier 1: Canonical}}
			                     & 16-bit PCM                                                 & $\le 48$\,kHz          & 72.30\%      & 71.53\%              & 61.97\%           & 64.24\%           & 64.31\%              & 63.10\%                     & $\mathbf{54.46 \pm 1.55}$\% \\ \cmidrule(lr){2-10}
			                     & 24-bit PCM                                                 & $> 48$\,kHz            & 93.38\%      & 89.48\%              & 73.30\%           & 74.59\%           & 74.55\%              & 68.16\%                     & $\mathbf{54.97 \pm 0.24}$\% \\ \midrule
			\multirow{4}{*}{\textbf{Tier 2: Stratified Music}}
			                     & 16-bit PCM                                                 & $\le 48$\,kHz          & 93.73\%      & 89.64\%              & 78.59\%           & 79.12\%           & 79.08\%              & 82.02\%                     & $\mathbf{70.71 \pm 1.23}$\% \\ \cmidrule(lr){2-10}
			                     & \multirow{2}{*}{24-bit PCM}                                & $\le 48$\,kHz          & 94.27\%      & 89.83\%              & 77.29\%           & 78.20\%           & 78.20\%              & 80.52\%                     & $\mathbf{69.65 \pm 5.67}$\% \\ \cmidrule(lr){3-10}
			                     &                                                            & $> 48$\,kHz            & 89.33\%      & 88.46\%              & 71.58\%           & 74.84\%           & 74.94\%              & 71.27\%                     & $\mathbf{60.15 \pm 2.03}$\% \\ \cmidrule(lr){2-10}
			                     & \multicolumn{2}{l}{\textit{Tier 2 Macro-Mean (40 Tracks)}} & 92.25\%                & 89.25\%      & 76.01\%              & 77.53\%           & 77.54\%           & 78.11\%              & $\mathbf{66.91 \pm 1.67}$\%                               \\ \bottomrule
		\end{tabular}%
	}
\end{table*}

\end{document}
}